\documentclass[oneside]{amsart}
\usepackage{color,pstricks,caption}   

\newcommand{\M}{\mathcal M}

\renewcommand\vert{\,|\,}

\newcommand{\wt}{{wt}}
\newcommand{\wh}{\widehat{wt}}
\newcommand{\unwt}{{{\hat{w}}^{\circ}}}

\newcommand{\halfh}{\hat h}
\newcommand{\hh}{\hat h}
\newcommand{\hcut}{h^{\mathrm{cut}}}
\newcommand{\hlac}{\hat h^{\mathrm{cut}}}
\newcommand{\vsymbol}{w}   
\newcommand{\vword}[1]{\vsymbol(#1)}
\newcommand{\osymbol}{\pi}
\newcommand{\oword}[1]{\osymbol(#1)}
\newcommand{\osymbolh}{\hat\pi}
\newcommand{\owordh}[1]{\osymbolh(#1)}

\newcommand{\PP}{{\mathcal P}}
\newcommand{\HH}{{\mathcal H}}
\newcommand{\BB}{{\mathcal B}}

\def\HH{\mathcal{H}}
\def\PP{\mathcal{P}}

\def\ch{\hat{c}}
\def\chs{\hat{c}^{*}}
\def\dh{\hat{d}}
\def\cb{c}
\def\cbs{c^{*}}
\def\db{d}

\let\rw\rightarrow
\let\y\infty

\def\ZZ{{\mathbb{Z}}}
\def\ZZph{{\mathbb{Z}+\tfrac12}}

\begin{document}
\title[Path bijection for $\M(p,2p+1)$: \today]{A bijection between
              paths for the $\M(p,2p+1)$ minimal model Virasoro characters}
\thanks{\today}

\author{Olivier Blondeau-Fournier}
\address{D\'epartement de physique, de g\'enie physique et
d'optique, Universit\'e Laval,  Qu\'ebec, Canada, G1K 7P4.}
\email{olivier.b-fournier.1@ulaval.ca}

\author{Pierre Mathieu}
\address{D\'epartement de physique, de g\'enie physique et
d'optique, Universit\'e Laval,  Qu\'ebec, Canada, G1K 7P4.}
\email{pmathieu@phy.ulaval.ca}

\author{Trevor Welsh}
\address{
Department of Physics,
University of Toronto,
Ontario,
Canada,
M5S 1A7.}
\email{trevor.welsh@utoronto.ca} 


\begin{abstract}
The states in the irreducible modules of the minimal models
can be represented by infinite lattice paths arising from
consideration of the corresponding RSOS statistical models.
For the $\M(p,2p+1)$ models, a completely different path
representation has been found recently, this one
on a half-integer
lattice; it has no known underlying statistical-model interpretation.
The correctness of this alternative representation has not yet been
demonstrated, even at the level of the generating functions,
since the resulting fermionic characters differ from the known ones.
This gap is filled here, with the presentation of two
versions of a bijection between the two path representations
of the $\M(p,2p+1)$ states.

In addition, a half-lattice path representation for the
$\M(p+1,2p+1)$ models is stated, and other generalisations suggested.
\end{abstract}

\maketitle

\let\a\alpha
\let\n\noindent

\section{Introduction}


The corner-transfer matrix is a powerful tool for expressing the
local state probabilities of the order variable in terms of
weighted sums of one-dimensional configurations \cite{Ba}.
For the restricted-solid-on-solid (RSOS) models \cite{ABF,FB}
in regime III
(in the infinite length limit),  
each weighted configuration sum
turns out to be the character of an  
irreducible module of the corresponding minimal model \cite{Kyoto,Mel}.
Each configuration is the specification of
the order variable's value at each position, with fixed extremities,
and the differences between neighbouring values constrained.
When dressed with edges linking adjacent points,
it becomes a lattice path.
Every state in the corresponding conformal theory is thus
represented by a particular lattice path \cite{OleJS,FLPW}.

This path description of the states has proved to be a royal road
for the construction of the fermionic characters \cite{Melzfermionic},
either using direct methods \cite{OleJS,PMprsos},
or recursive ones \cite{FLPW,FW-kyoto,TWfe}.
For the special class of minimal models $\M(p,2p+1)$,
an alternative path representation was found in \cite{PMnpd},
and this led to novel fermionic expressions.
These paths, however, were not deduced from a statistical model
whose scaling limit was known to be the $\M(p,2p+1)$ minimal model.
Moreover, their generating functions have not, hitherto,
been proved to be equivalent to either the bosonic expressions
\cite{Rocha-Caridi} or the usual fermionic expressions
\cite{Melzfermionic,FQ,FW-kyoto,TWfe,PMprsos}
for these characters
(although the equivalence has been checked to high order,
and was supported by an asymptotic analysis).
Albeit formally ad hoc, the discovery of this new path representation
relied on heuristic considerations which we briefly recall.

Like their scaling limit, the RSOS models are parameterised by two
relatively prime positive integers $p,\,p'$ with
$p'>p\geq 2$.%
\footnote{For the RSOS models/paths, we follow the notation used in
  \cite{FB,FLPW,TWfe,FW-kyoto}.
  Consequently, the corresponding notation for the minimal models
  differs from that of \cite{CFT,PMprsos} 
  by the interchange of $p$ and $p'$.
  This interchange is of course irrelevant, the physical characteristics,
  the conformal dimensions
  $$h_{r,s}= \frac{[r\,{\rm max}(p,p')-s\,{\rm min}(p,p')]^2-(p-p')^2}{4 p p'}
  $$
  of the highest-weight states in modules labelled by $(r,s)$ for 
  $1\leq r<{\rm min}(p,p')$ and
  $1\leq s<{\rm max}(p,p')$,
  remain unchanged.
}
The corresponding paths are sequences of NE and SE edges restricted to
the interior of an infinite strip of vertical {width} 
$p'-2$
(this definition is made more precise in Section \ref{Sec:FBpaths}).
The weight of each path
is the sum of the weights of its vertices.
The weight of a vertex depends
not only upon its shape, but also
upon the two parameters $p$ and $p'$ \cite{FB}.
In the unitary cases (those cases where $p'=p+1$),
a drastic simplification occurs in
that the peaks and the valleys of a path
do not contribute to its 
weight, while every other vertex contributes
$x/2$, where $x$ is its 
horizontal position.
On the other hand, weighting the path in the dual way
(assigning 
weight $x/2$ to the valleys and peaks and 0 to all other vertices)
leads to the character of the $\mathcal Z_{p-1}$
parafermionic models \cite{BMlmp,OleJS,Kyoto,FWmel}.
These cases were then contrasted with
the paths describing the states of the
graded versions
of these parafermionic models -- equivalent to
the cosets $\widehat{osp}(1,2)_{p-1} /\hat u(1)$ \cite{CRS,JMgra}.
These latter paths reside
on a half-integer lattice with only peaks and valleys
contributing to the weight, each to the amount $x/2$, and with the
special constraint that peaks are forced to occur at integer heights.%
\footnote{The usual parafermionic states have two path representations,
  bijectively related in \cite{JMpath}
  (see also \cite{PMjmp} and references therein).
  This is also true for the graded ones \cite[\S5]{PMnpd}.
  We are referring here to the RSOS-type path representation
  (with no horizontal edges).}
Seeking their dual versions led to the new
path description of the $\M(p,2p+1)$ models \cite{PMnpd}.

The demonstration of the correctness of this new path description
of the $\M(p,2p+1)$ states is the main subject of the present work.
This is achieved by exhibiting a bijection between these paths
and the RSOS paths.
In fact, two equivalent but quite different-looking versions
of this bijection are presented.

After defining the two types of path in Section \ref{Sec:2paths}, the
first version of the bijection is presented in Section \ref{Sec:CoBi}.
It relies on techniques developed in \cite{FLPW,TWfe}.
It essentially amounts to the deconstruction of an RSOS path,
followed by a corresponding construction of a half-lattice path.
The second version of the bijection is presented in
Section \ref{Sec:OperatorBij}.
It relies on an encoding of the vertex words for RSOS paths
introduced in \cite{TWpa}, and an analogous construction for
the half-lattice paths that is
a modification of that given in \cite{PMnlob}.
%
In this approach, each path, with given extremity conditions,
is constructed from a sequence of operators acting on 
a suitable ground-state path.
The bijection is expressed as a rule for transforming
between sequences of operators that describe the two paths.
Although the two versions of the bijection were originally obtained
independently, we show in Section \ref{Sec:BijEquiv} that the
second is, in fact, a transformation of the first,
thereby immediately obtaining its proof.

\section{Defining the two types of paths}\label{Sec:2paths}

\subsection{$\a$-Lattice paths}\label{Sec:Lattice}

An infinite length $\a$-lattice path $h$ is {a} sequence
$(h_0,h_\a,h_{2\a},\ldots)$ for which $|h_{x+\a}-h_{x}|=\a$ for
$x\in{ \a}\ZZ_{\ge0}$.
The path $h$ is said to be $(f,g)$-restricted if
$f\le h_x\le g$ for all $x\ge0$ and $b$-tailed
if there exists $L\ge0$ such that $h_x\in\{b,b-\a\}$ for all $x\ge L$.
In this work, we consider two cases:
$\a=1$ (integer lattice paths) and $\a=1/2$
(half-lattice paths).

By linking the points $(0,h_0)$, $(\a,h_\a)$, $(2\a,h_{2\a}),\ldots$
on the $x$-$y$ plane we obtain a graph that conveniently depicts the path:
we refer to this graph as the path picture.
For each $x>0$, the position $(x,h_x)$ on the
path picture of $h$ is referred to as a vertex.
The shape of the vertex is either a peak, valley,
straight-up or straight-down depending on whether the edges
on the two sides of the vertex are NE-SE, SE-NE, NE-NE or SE-SE,
respectively.


On occasion, it will be necessary to refer also to the position
$(0,h_0)$ as a vertex, and to have its shape determined.
When required, this is done by specifying the value of
$h_{-\a}$ to be one of the two values $h_0\pm\a$.
In effect, this defines the direction of a path pre-segment.

\subsection{RSOS paths and the $\mathcal {M}(p,p')$ models}
\label{Sec:FBpaths}

\let\rw\rightarrow
\let\y\infty

The set $\PP^{p,p'}_{a,b}$ of RSOS
paths is defined to be the set of infinite%
\footnote{These paths were originally defined for finite
length in \cite{FB} and studied extensively in \cite{FLPW,FW-kyoto,TWfe}.
The collection of finite paths (with specific boundary conditions)
provides a finitized version of the Virasoro characters \cite{Mel}.
The finitization parameter enables the derivation of useful
recurrence relations for the path generating functions.
In particular, one can define the notion of dual finitized characters,
which are obtained via the transformation
$q\rw1/q$ \cite{BMlmp,OleJS,FLPW, PMnpd}.
}
length integer lattice paths $h$ that are $(1,p'-1)$-restricted,
$b$-tailed, 
with $h_0=a$.
Each RSOS
path $h\in\PP^{p,p'}_{a,b}$ is assigned a
weight $\wt(h)$ which we now define.%
\footnote{The definition given here is that of \cite{FLPW}.
  It differs considerably from that originally given {in} \cite{FB}.
  In fact, although this is not obvious, the weighting
  of the paths $\PP^{p,p'}_{a,b}$ defined in
  \cite{FB} differs from that of \cite{FLPW} by an overall
  constant depending upon $a$ and $b$.}

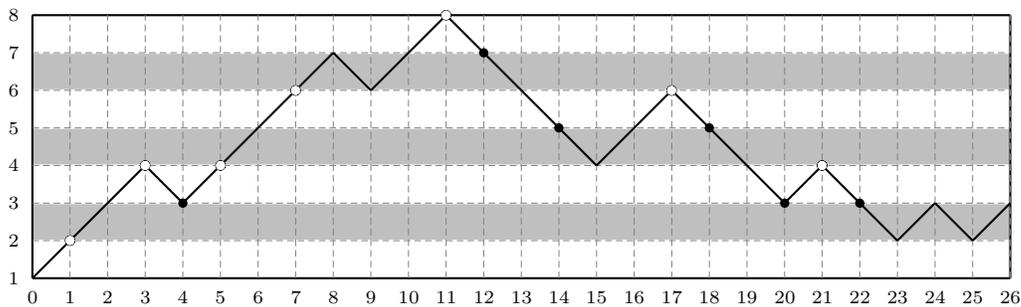
\begin{figure}[ht]
\caption{{\footnotesize A path $h\in \PP^{4,9}_{1,3}$.
The unfilled circles indicate up-scoring vertices, which have weight $u_x$,
and the filled circles indicate down-scoring vertices, which have
weight $v_x$, both values being defined in (\ref{Eq:ScoreWts}).}}
\vskip-1.2cm
\label{TypicalScoringFig}
\begin{center}
\begin{pspicture}(1,0)(13,5.5)

{
\psset{linestyle=none}
\psset{fillstyle=solid}
\psset{fillcolor=lightgray}
\psframe(0.5,1)(13.5, 1.5)
\psframe(0.5,2)(13.5, 2.5)
\psframe(0.5,3)(13.5, 3.5)
}

\psset{linestyle=solid}
\psline{-}(0.5,0.5)(13.5,0.5)
\psline{-}(0.5,4)(13.5,4)

\psline{-}(0.5,0.5)(0.5,4)
\psline{-}(13.5,0.5)(13.5,4)

{
\psset{linewidth=0.25pt,linestyle=dashed, dash=2.5pt 1.5pt,linecolor=gray}

\psline{-}(0.5,1.0)(13.5,1.0)
\psline{-}(0.5,1.5)(13.5,1.5)
\psline{-}(0.5,2.0)(13.5,2.0)
\psline{-}(0.5,2.5)(13.5,2.5)
\psline{-}(0.5,3)(13.5,3)
\psline{-}(0.5,3.5)(13.5,3.5)

\psline{-}(1.0,0.5)(1.0,4.0) \psline{-}(1.5,0.5)(1.5,4.0) 
\psline{-}(2.0,0.5)(2.0,4.0) \psline{-}(2.5,0.5)(2.5,4.0) \psline{-}(3.0,0.5)(3.0,4.0) 
\psline{-}(3.5,0.5)(3.5,4.0) \psline{-}(4.0,0.5)(4.0,4.0) \psline{-}(4.5,0.5)(4.5,4.0) 
\psline{-}(5.0,0.5)(5.0,4.0) \psline{-}(5.5,0.5)(5.5,4.0) \psline{-}(6.0,0.5)(6.0,4.0) 
\psline{-}(6.5,0.5)(6.5,4.0) \psline{-}(7.0,0.5)(7.0,4.0) \psline{-}(7.5,0.5)(7.5,4.0) 
\psline{-}(8.0,0.5)(8.0,4.0) \psline{-}(8.5,0.5)(8.5,4.0) \psline{-}(9.0,0.5)(9.0,4.0) 
\psline{-}(9.5,0.5)(9.5,4.0) \psline{-}(10.0,0.5)(10.0,4.0) \psline{-}(10.5,0.5)(10.5,4.0) 
\psline{-}(11.0,0.5)(11.0,4.0) \psline{-}(11.5,0.5)(11.5,4.0) \psline{-}(12.0,0.5)(12.0,4.0) 
\psline{-}(12.5,0.5)(12.5,4.0) \psline{-}(13.0,0.5)(13.0,4.0)  \psline{-}(13.5,0.5)(13.5,4.0)
}

{
\rput(0.25,0.5){{\scriptsize $1$}}
\rput(0.25,1.0){{\scriptsize $2$}} \rput(0.25,1.5){{\scriptsize $3$}}
\rput(0.25,2.0){{\scriptsize $4$}} \rput(0.25,2.5){{\scriptsize $5$}}
\rput(0.25,3.0){{\scriptsize $6$}} \rput(0.25,3.5){{\scriptsize $7$}}
\rput(0.25,4.0){{\scriptsize $8$}}

\rput(0.5,0.25){{\scriptsize $0$}}
\rput(1.0,0.25){{\scriptsize $1$}} \rput(1.5,0.25){{\scriptsize $2$}} \rput(2.0,0.25){{\scriptsize $3$}}
\rput(2.5,0.25){{\scriptsize $4$}} \rput(3.0,0.25){{\scriptsize $5$}} \rput(3.5,0.25){{\scriptsize $6$}}
\rput(4.0,0.25){{\scriptsize $7$}} \rput(4.5,0.25){{\scriptsize $8$}} \rput(5.0,0.25){{\scriptsize $9$}} 
\rput(5.5,0.25){{\scriptsize $10$}} \rput(6.0,0.25){{\scriptsize $11$}} \rput(6.5,0.25){{\scriptsize $12$}}
\rput(7.0,0.25){{\scriptsize $13$}} \rput(7.5,0.25){{\scriptsize $14$}} \rput(8.0,0.25){{\scriptsize $15$}}
\rput(8.5,0.25){{\scriptsize $16$}} \rput(9.0,0.25){{\scriptsize $17$}} \rput(9.5,0.25){{\scriptsize $18$}}
\rput(10.0,0.25){{\scriptsize $19$}} \rput(10.5,0.25){{\scriptsize $20$}} \rput(11.0,0.25){{\scriptsize $21$}}   
\rput(11.5,0.25){{\scriptsize $22$}} \rput(12.0,0.25){{\scriptsize $23$}} \rput(12.5,0.25){{\scriptsize $24$}}
\rput(13.0,0.25){{\scriptsize $25$}} \rput(13.5,0.25){{\scriptsize $26$}} 
}

{
\psset{linestyle=solid}

\psline{-}(.5,.5)(2.0,2.0) 
\psline{-}(2.0,2.0)(2.5,1.5) \psline{-}(2.5,1.5)(4.5,3.5)
\psline{-}(4.5,3.5)(5.0,3.0) \psline{-}(5.0,3.0)(5.5,3.5)
\psline{-}(5.5,3.5)(6.0,4.0) \psline{-}(6.0,4.0)(8.0,2.0)
\psline{-}(8.0,2.0)(9.0,3.0) \psline{-}(9.0,3.0)(10.5,1.5)
\psline{-}(10.5,1.5)(11.0,2.0) \psline{-}(11.0,2.0)(12.0,1.0)
\psline{-}(12.0,1.0)(12.5,1.5) \psline{-}(12.5,1.5)(13.0,1.0)
\psline{-}(13.0,1.0)(13.5,1.5)
}
{
\psset{dotsize=3.5pt}
\psdots(2.5,1.5)(6.5,3.5)(7.5,2.5)(9.5,2.5)(10.5,1.5)(11.5,1.5)
\psset{fillcolor=white}
\psset{dotsize=4pt}\psset{dotstyle=o}
\psdots(1,1)(2,2)(3,2)(4,3)(6,4)(9,3)(11,2)
}
\end{pspicture}
\end{center}
\end{figure}

For $1\le k\le p'-1$, we refer to the region of the $x$-$y$ plane
between $y=k$ and $y=k+1$ as the $k$th band.
Thus, in the path picture, the paths from $\PP^{p,p'}_{a,b}$
lie in the $p'-2$ bands between $y=1$ and $y=p'-1$.
For $1\le r<p$, we shade the $\lfloor rp'/p\rfloor$th band and refer
to it as the $r$th dark band.
Bands that are not dark are referred to as light bands.%
\footnote{In \cite{FLPW,FW-kyoto,TWfe}, dark and light bands were
  referred to as odd and even bands respectively.}
%
The band structure for the case $(p,p')=(4,9)$, along
with a typical path from $\PP^{4,9}_{1,3}$,
is shown in Fig.~\ref{TypicalScoringFig}.
This band pattern is typical of all models with $p'=2p+1$: there are
$p-1$ dark bands separated from each other by a single light one.

Those vertices which either are straight with the right edge in a dark
band, or are not straight with the right edge in a light band,
are referred to as \emph{scoring} vertices.
All other vertices are referred to as \emph{non-scoring} vertices.
Each scoring vertex is said to be \emph{up-scoring} or
\emph{down-scoring} depending on whether the left edge is
up or down respectively.
In the path picture, we (often) highlight the up-scoring vertices
with an unfilled circle, and the down-scoring vertices with a
filled {circle}. 
This is done for the path of Fig.~\ref{TypicalScoringFig}.

After setting%
\footnote{The point $(x,h_x)$ has coordinates $(u_x,v_x)$
  in the system where the axes are inclined at $45^o$, and the
  origin is at the path startpoint.}
\begin{equation}\label{Eq:ScoreWts}
u_x=\frac12 (x-h_x+a)\qquad
v_x=\frac12 (x+h_x-a),
\end{equation}
the weight $\wt(h)$ of a path $h$ is defined to be:
\begin{equation}\label{Eq:WtDef}
\wt(h)=\sum_{{x>0}}w_x,   
\end{equation}
where we define
\begin{equation}
w_x=
\begin{cases}
u_x&\text{if $(x,h_x)$ is up-scoring;}\\
v_x&\text{if $(x,h_x)$ is down-scoring;}\\
0&\text{if $(x,h_x)$ is non-scoring.}
\end{cases}
\end{equation}

\n For the path $h$ depicted in Fig.~\ref{TypicalScoringFig},
we obtain 
\begin{equation}\wt(h)=0+0+3+1+1+2+9+9+6+11+11+9+12=74.
\end{equation}


The tail condition of an RSOS path $h$ indicates that after a certain
position, the path forever oscillates within a single band.
The definition \eqref{Eq:WtDef} then implies that $\wt(h)$ is finite
or infinite depending on whether this band is dark or light respectively.
Below, we restrict to those cases where that band is dark.


The paths in the set $\PP^{p,p'}_{a,b}$ provide a combinatorial
description of the states in the irreducible module
of the minimal model $\M(p,p')$ that is labelled by $(r,s)$, 
where the extremity points $a$ and $b$ are related to $r$ and $s$ by
\begin{equation}\label{rsab}
a=s\quad \text{and}\quad b=\lfloor rp'/p\rfloor+1.
\end{equation}
(Thereupon, the heights $b$ and $b-1$ straddle
the $r$th dark band of the path picture).
The generating function of the set $\PP^{p,p'}_{a,b}$ of paths is 
then the Virasoro character $\chi^{p,p'}_{r,s}$
\cite{FLPW}:
\begin{equation}\label{Eq:GFvirasoro}
\sum_{h\in\PP^{p,p'}_{a,b}} q^{\wt(h)}
=\chi^{p,p'}_{r,s}(q).
\end{equation}


\subsection{$\M(p,2p+1)$ characters and half-lattice paths}
\label{Sec:Halfpath}

The states in the irreducible modules of the $\M(p,2p+1)$ models
turn out to have a path representation alternative to that
specified above. This description was first given in \cite{PMnpd}.

For $p\ge2$ and $\hat a,\hat b\in\ZZ$,
define $\HH^p_{\hat a,\hat b}$ to be the set of all
half-lattice paths $\halfh$ that are $(0,p-1)$-restricted,
$\hat b$-tailed, with $\halfh_0=\hat a$, and 
the \emph{additional} restriction that if
$\halfh_x=\halfh_{x+1}\in\ZZ$, then $\halfh_{x+1/2}=\halfh_x-1/2$.
The additional restriction here implies that peaks can only
occur at integer heights.

For each path $\halfh\in\HH^p_{\hat a,\hat b}$, we specify
the shape of the vertex at its startpoint $(0,\hat a)$
by adopting the convention that $\halfh_{-1/2}=\hat a-1/2$
(we do this even if $\hat a=0$).
%


The \emph{unnormalised} weight $\unwt(\halfh)$ of a path 
$\halfh\in\HH^p_{\hat a,\hat b}$ is defined to be half the sum of
those $x\in\tfrac12{\ZZ_{\ge0}}$ for which $(x,\halfh_x)$ is a
straight-vertex:
\begin{equation}\label{Eq:StraightWt}
\unwt(\halfh)
=\frac12\sum_{\begin{subarray}{c} x\in\frac12{\ZZ_{\ge0}}\\
                                  \halfh_{x-1/2}\ne \halfh_{x+1/2}
              \end{subarray}} x.
\end{equation}

Define the \emph{ground-state path} 
$\halfh^{\mathrm{gs}}\in\HH^p_{\hat a,\hat b}$ to be that
path which has minimal weight amongst all the elements
of $\HH^p_{\hat a,\hat b}$.
It is easily seen that this path has the following shape:
an oscillating part
having
$\halfh^{\mathrm{gs}}_x=\hat b$ and
$\halfh^{\mathrm{gs}}_{x+1/2}=\hat b-1/2$ for
$x\in\ZZ_{\ge | \hat a- \hat b| }$,
%
%
preceded by an initial straight line {having} $\halfh^{\mathrm{gs}}_x
  =(\hat a, \hat a \pm 1/2, \hat a \pm 1, \ldots , \hat b \mp 1/2)$
for $x=(0,1/2,1,\ldots,|\hat a - \hat b | -1/2)$, where the
upper signs apply when $\hat b \ge \hat a$ and the
lower signs apply when $\hat b < \hat a$.

\begin{figure}[ht]
\caption{{\footnotesize A typical path
          ${\halfh}\in \mathcal{H}_{0,1}^{4}$.
          The black dots indicate the vertices that contribute to the weight,
          with that at position $(x,\halfh_x)$ contributing $x/2$.
          }}
\label{che_h}
\begin{center}
\begin{pspicture}(2.3,-0.5)(12.5,2.3)
{\psset{yunit=0.32cm,xunit=0.32cm,linewidth=.6pt}
{
\psset{linewidth=0.25pt,linestyle=dashed, dash=2.5pt 1.5pt,linecolor=gray}
\psline{-}(0,1)(46,1.0)
\psline{-}(0,2)(46,2.0)
\psline{-}(0,3)(46,3.0)
\psline{-}(0,4)(46,4.0)
\psline{-}(0,5)(46,5.0)
\psline{-}(0,6)(46,6.0)
\psline{-}(1,0)(1,6) \psline{-}(2,0)(2,6) \psline{-}(3,0)(3,6)
\psline{-}(4,0)(4,6) \psline{-}(5,0)(5,6) \psline{-}(6,0)(6,6)
\psline{-}(7,0)(7,6) \psline{-}(8,0)(8,6) \psline{-}(9,0)(9,6) 
\psline{-}(10,0)(10,6) \psline{-}(11,0)(11,6) \psline{-}(12,0)(12,6)
\psline{-}(13,0)(13,6) \psline{-}(14,0)(14,6) \psline{-}(15,0)(15,6) 
\psline{-}(16,0)(16,6) \psline{-}(17,0)(17,6) \psline{-}(18,0)(18,6)
\psline{-}(19,0)(19,6) \psline{-}(20,0)(20,6) \psline{-}(21,0)(21,6) 
\psline{-}(22,0)(22,6) \psline{-}(23,0)(23,6) \psline{-}(24,0)(24,6) \psline{-}(25,0)(25,6)
\psline{-}(26,0)(26,6) \psline{-}(27,0)(27,6) \psline{-}(28,0)(28,6) 
\psline{-}(29,0)(29,6) \psline{-}(30,0)(30,6) \psline{-}(31,0)(31,6)
\psline{-}(32,0)(32,6) \psline{-}(33,0)(33,6) \psline{-}(34,0)(34,6) 
\psline{-}(35,0)(35,6) \psline{-}(36,0)(36,6) \psline{-}(37,0)(37,6)
\psline{-}(38,0)(38,6) \psline{-}(39,0)(39,6) \psline{-}(40,0)(40,6) 
\psline{-}(41,0)(41,6) \psline{-}(42,0)(42,6) \psline{-}(43,0)(43,6)
\psline{-}(44,0)(44,6) \psline{-}(45,0)(45,6) \psline{-}(46,0)(46,6)
}
{
\psline{-}(0,0)(0,6)
\psline{->}(0,0)(46,0)
\psline(0,1)(0.2,1)
\psline(0,2)(0.2,2)
\psline(0,3)(0.2,3)
\psline(0,4)(0.2,4)
\psline(0,5)(0.2,5)
\psline(0,6)(0.5,6)

\psline(1,0)(1,.2)     \psline(2,0)(2,.2)        \psline(3,0)(3,.2)
\psline(4,0)(4,.2)     \psline(5,0)(5,.2)        \psline(6,0)(6,.2)
\psline(7,0)(7,.2)     \psline(8,0)(8,.2)        \psline(9,0)(9,.2)
\psline(10,0)(10,.2)     \psline(11,0)(11,.2)        \psline(12,0)(12,.2)
\psline(13,0)(13,.2)     \psline(14,0)(14,.2)        \psline(15,0)(15,.2)
\psline(16,0)(16,.2)     \psline(17,0)(17,.2)        \psline(18,0)(18,.2)
\psline(19,0)(19,.2)     \psline(20,0)(20,.2)        \psline(21,0)(21,.2)
\psline(22,0)(22,.2)    \psline(23,0)(23,.2)    \psline(24,0)(24,.2)    
\psline(25,0)(25,.2)   \psline(26,0)(26,.2)    \psline(27,0)(27,.2)   
\psline(28,0)(28,.2)   \psline(29,0)(29,.2)    \psline(30,0)(30,.2)
\psline(31,0)(31,.2)   \psline(32,0)(32,.2)    \psline(33,0)(33,.2)
\psline(34,0)(34,.2)   \psline(35,0)(35,.2)    \psline(36,0)(36,.2)
\psline(37,0)(37,.2)   \psline(38,0)(38,.2)    \psline(39,0)(39,.2)
\psline(40,0)(40,.2)   \psline(41,0)(41,.2)    \psline(42,0)(42,.2)
\psline(43,0)(43,.2)   \psline(44,0)(44,.2)    \psline(45,0)(45,.2)
}

{
\rput(-.5,0){\scriptsize 0}
\rput(-.5,2){\scriptsize 1}
\rput(-.5,4){\scriptsize 2}
\rput(-.5,6){\scriptsize 3}

\rput(0,-0.5){\scriptsize 0}
\rput(2,-0.5){\scriptsize 1}    \rput(4,-0.5){\scriptsize 2}
\rput(6,-0.5){\scriptsize 3}    \rput(8,-0.5){\scriptsize 4}
\rput(10,-0.5){\scriptsize 5}   \rput(12,-0.5){\scriptsize 6}
\rput(14,-0.5){\scriptsize 7}   \rput(16,-0.5){\scriptsize 8}
\rput(18,-0.5){\scriptsize 9}   \rput(20,-0.5){\scriptsize 10} 
\rput(22,-0.5){\scriptsize 11}  \rput(24,-0.5){\scriptsize 12}
\rput(26,-0.5){\scriptsize 13}  \rput(28,-0.5){\scriptsize 14}
\rput(30,-0.5){\scriptsize 15}  \rput(32,-0.5){\scriptsize 16}
\rput(34,-0.5){\scriptsize 17}  \rput(36,-0.5){\scriptsize 18}
\rput(38,-0.5){\scriptsize 19}  \rput(40,-0.5){\scriptsize 20}
\rput(42,-0.5){\scriptsize 21}  \rput(44,-0.5){\scriptsize 22}
}

{
\psline(0,0)(6,6)
\psline(6,6)(7,5)
\psline(7,5)(8,6)
\psline(8,6)(12,2)
\psline(12,2)(14,4)
\psline(14,4)(17,1)
\psline(17,1)(18,2)
\psline(18,2)(19,1)
\psline(19,1)(20,2)
\psline(20,2)(21,1)
\psline(21,1)(22,2)
\psline(22,2)(23,1)
\psline(23,1)(24,2)
\psline(24,2)(26,0)
\psline(26,0)(28,2)
\psline(28,2)(29,1)
\psline(29,1)(30,2)
\psline(30,2)(31,1)
\psline(31,1)(32,2)
\psline(32,2)(33,1)
\psline(33,1)(34,2)
\psline(34,2)(35,1)
\psline(35,1)(36,0)
\psline(36,0)(38,2)
\psline(38,2)(39,1)
\psline(39,1)(40,2)
\psline(40,2)(41,1)
\psline(41,1)(42,0)
\psline(42,0)(44,2)
\psline(44,2)(45,1)\psline(45,1)(46,2)
}
{
\psset{dotsize=2.5pt}
\psdots(0,0)(1,1)(2,2)(3,3)(4,4)(5,5)(9,5)(10,4)(11,3)(13,3)%
(15,3)(16,2)(25,1)(27,1)(35,1)(37,1)(41,1)(43,1)

}
}
\end{pspicture}
\end{center}
\end{figure}
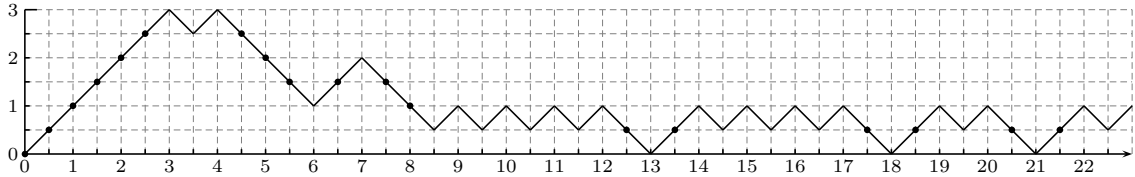

The weight $\wh(\halfh)$ of each $\halfh\in\HH^p_{\hat a,\hat b}$ is
then defined by
\begin{equation}\label{Eq:HalfWt}
\wh(\halfh)=\unwt(\halfh)-\unwt(\halfh^{\mathrm{gs}}).
\end{equation}
For instance, in the path $\halfh\in\HH^4_{0,1}$ of Fig.~\ref{che_h},
we have used dots {to} indicate the vertices that contribute
to \eqref{Eq:StraightWt}.
The corresponding ground-state path
$\halfh^{\mathrm{gs}}\in\HH^4_{0,1}$ has
$\halfh^{\mathrm{gs}}_x=1$ for $x\in\ZZ_{>0}$
and
$\halfh^{\mathrm{gs}}_x=1/2$ for $x\in\ZZ_{\ge0}+1/2$.
Therefore, from \eqref{Eq:HalfWt}, we obtain:
\begin{equation}
\begin{split}
\wh(\halfh) &= \tfrac{1}{2} \left(0 + \tfrac12+ 1+\tfrac{3}{2}
 + 2 + \tfrac52 + \tfrac92 + 5 + \tfrac{11}{2} + \tfrac{13}{2}
+\tfrac{15}{2}+ 8
\right. \\
&\hskip 39mm
\left.
+\tfrac{25}{2} + \tfrac{27}{2} +\tfrac{35}{2} + \tfrac{37}{2}
+\tfrac{41}{2} + \tfrac{43}{2} \right)
-\tfrac{1}{2} \left(0 + \tfrac12 \right)
= 74.
\end{split}
\end{equation}

%
%
In what follows, we make use of the following trick to calculate
$\wh(\halfh)$ directly from the path picture of
$\halfh\in\HH^p_{\hat a,\hat b}$.
First note that the minimal weight path
$\halfh^{\mathrm{gs}}\in\HH^p_{\hat a,\hat b}$
extends between heights $\hat a$ and $\hat b$
in its first $2e$ (half-integer) steps, where $e=|\hat a-\hat b|$.
An alternative to \eqref{Eq:HalfWt} for obtaining $\wh(\halfh)$,
is to extend $\halfh$
to the left by $2e$ steps, in such a way that $\halfh_{-e}=\hat b$
(overriding the above convention for $\halfh_{-1/2}$).
The renormalised weight $\wh(\halfh)$ is then obtained by summing the
$x$-coordinates of all the straight vertices of this extended path,
beginning with its first vertex at $(-e,\hat b)$ whose nature
is specified by setting $\halfh_{-e-1/2}=\hat b-1/2$,
and dividing by 2.

\begin{figure}[ht]
\caption{{\footnotesize A path ${\halfh}\in\HH^5_{3,1}$ (solid) and the
         corresponding ground state ${\halfh}^{\text{gs}}$ (dotted).}}
\vskip-0.5cm
\label{Fig:UnNorm}
\begin{center}
\begin{pspicture}(3.0,-0.5)(3.5,3.3)
{\psset{yunit=0.33cm,xunit=0.33cm,linewidth=.6pt}

{
\psset{linewidth=0.25pt,linestyle=dashed, dash=2.5pt 1.5pt,linecolor=gray}

\psline{-}(0,1)(20,1.0)
\psline{-}(0,2)(20,2.0)
\psline{-}(0,3)(20,3.0)
\psline{-}(0,4)(20,4.0)
\psline{-}(0,5)(20,5.0)
\psline{-}(0,6)(20,6.0)
\psline{-}(0,7)(20,7.0)
\psline{-}(0,8)(20,8.0)

\psline{-}(1,0)(1,8) \psline{-}(2,0)(2,8) \psline{-}(3,0)(3,8) 
\psline{-}(4,0)(4,8) \psline{-}(5,0)(5,8) \psline{-}(6,0)(6,8)
\psline{-}(7,0)(7,8) \psline{-}(8,0)(8,8) \psline{-}(9,0)(9,8) 
\psline{-}(10,0)(10,8) \psline{-}(11,0)(11,8) \psline{-}(12,0)(12,8)
\psline{-}(13,0)(13,8) \psline{-}(14,0)(14,8) \psline{-}(15,0)(15,8) 
\psline{-}(16,0)(16,8) \psline{-}(17,0)(17,8) \psline{-}(18,0)(18,8)
\psline{-}(19,0)(19,8) \psline{-}(20,0)(20,8) 
}


\psline{-}(0,0)(0,8)
\psline{->}(0,0)(20,0)

\psline(0,1)(0.2,1)
\psline(0,2)(0.2,2)
\psline(0,3)(0.2,3)
\psline(0,4)(0.2,4)
\psline(0,5)(0.2,5)
\psline(0,6)(0.2,6)
\psline(0,7)(0.2,7)
\psline(0,8)(0.5,8)

\psline(1,0)(1,.2)     \psline(2,0)(2,.2)        \psline(3,0)(3,.2)
\psline(4,0)(4,.2)     \psline(5,0)(5,.2)        \psline(6,0)(6,.2)
\psline(7,0)(7,.2)     \psline(8,0)(8,.2)        \psline(9,0)(9,.2)
\psline(10,0)(10,.2)     \psline(11,0)(11,.2)        \psline(12,0)(12,.2)
\psline(13,0)(13,.2)     \psline(14,0)(14,.2)        \psline(15,0)(15,.2)
\psline(16,0)(16,.2)     \psline(17,0)(17,.2)        \psline(18,0)(18,.2)
\psline(19,0)(19,.2)     


\rput(-.5,0){\scriptsize 0}
\rput(-.5,2){\scriptsize 1}
\rput(-.5,4){\scriptsize 2}
\rput(-.5,6){\scriptsize 3}
\rput(-.5,8){\scriptsize 4}

\rput(0,-0.5){\scriptsize 0}
\rput(2,-0.5){\scriptsize 1}    \rput(4,-0.5){\scriptsize 2}
\rput(6,-0.5){\scriptsize 3}    \rput(8,-0.5){\scriptsize 4}
\rput(10,-0.5){\scriptsize 5}   \rput(12,-0.5){\scriptsize 6}
\rput(14,-0.5){\scriptsize 7}   \rput(16,-0.5){\scriptsize 8}
\rput(18,-0.5){\scriptsize 9}   \rput(20,-0.5){\scriptsize 10}

\psline(0,6)(1,5)
\psline(1,5)(2,6)
\psline(2,6)(3,5)
\psline(3,5)(6,8)
\psline(6,8)(7,7)
\psline(7,7)(8,8)
\psline(8,8)(16,0)
\psline(16,0)(18,2)
\psline(18,2)(19,1)
\psline(19,1)(20,2)

{
\psset{linewidth=0.6pt,linestyle=dashed, dash=4pt 2pt,linecolor=black}

\psline{-}(1,5)(5,1)
\psline{-}(5,1)(6,2)
\psline{-}(6,2)(7,1)
\psline{-}(7,1)(8,2)
\psline{-}(8,2)(9,1)
\psline{-}(9,1)(10,2)
}
}
\end{pspicture}
\end{center}
\end{figure}

\begin{figure}[ht]
\caption{{\footnotesize The extension of the path {$\halfh$}
                        of Fig.~\ref{Fig:UnNorm}.}}
\vskip-0.5cm
\label{Fig:ReNorm}
\begin{center}
\begin{pspicture}(4.,-0.5)(5.0,3.3)
{\psset{yunit=0.33cm,xunit=0.33cm,linewidth=.6pt}

{
\psset{linewidth=0.25pt,linestyle=dashed, dash=2.5pt 1.5pt,linecolor=gray}

\psline{-}(0,1)(24,1.0)
\psline{-}(0,2)(24,2.0)
\psline{-}(0,3)(24,3.0)
\psline{-}(0,4)(24,4.0)
\psline{-}(0,5)(24,5.0)
\psline{-}(0,6)(24,6.0)
\psline{-}(0,7)(24,7.0)
\psline{-}(0,8)(24,8.0)

\psline{-}(1,0)(1,8) \psline{-}(2,0)(2,8) \psline{-}(3,0)(3,8) 
\psline{-}(0,0)(0,8) \psline{-}(5,0)(5,8) \psline{-}(6,0)(6,8)
\psline{-}(7,0)(7,8) \psline{-}(8,0)(8,8) \psline{-}(9,0)(9,8) 
\psline{-}(10,0)(10,8) \psline{-}(11,0)(11,8) \psline{-}(12,0)(12,8)
\psline{-}(13,0)(13,8) \psline{-}(14,0)(14,8) \psline{-}(15,0)(15,8) 
\psline{-}(16,0)(16,8) \psline{-}(17,0)(17,8) \psline{-}(18,0)(18,8)
\psline{-}(19,0)(19,8) \psline{-}(20,0)(20,8) \psline{-}(21,0)(21,8)
\psline{-}(22,0)(22,8) \psline{-}(23,0)(23,8) \psline{-}(24,0)(24,8)
}


\psline{-}(0,0)(0,8)
\psline{-}(4,0)(4,8)
\psline{->}(0,0)(24,0)

\psline(0,1)(0.2,1)
\psline(0,2)(0.2,2)
\psline(0,3)(0.2,3)
\psline(0,4)(0.2,4)
\psline(0,5)(0.2,5)
\psline(0,6)(0.2,6)
\psline(0,7)(0.2,7)
\psline(0,8)(0.5,8)

\psline(1,0)(1,.2)     \psline(2,0)(2,.2)        \psline(3,0)(3,.2)
\psline(4,0)(4,.2)     \psline(5,0)(5,.2)        \psline(6,0)(6,.2)
\psline(7,0)(7,.2)     \psline(8,0)(8,.2)        \psline(9,0)(9,.2)
\psline(10,0)(10,.2)     \psline(11,0)(11,.2)        \psline(12,0)(12,.2)
\psline(13,0)(13,.2)     \psline(14,0)(14,.2)        \psline(15,0)(15,.2)
\psline(16,0)(16,.2)     \psline(17,0)(17,.2)        \psline(18,0)(18,.2)
\psline(19,0)(19,.2)     


\rput(-.5,0){\scriptsize 0}
\rput(-.5,2){\scriptsize 1}
\rput(-.5,4){\scriptsize 2}
\rput(-.5,6){\scriptsize 3}
\rput(-.5,8){\scriptsize 4}

\rput(0,-0.5){\scriptsize -2}
\rput(2,-0.5){\scriptsize -1}    \rput(4,-0.5){\scriptsize 0}
\rput(6,-0.5){\scriptsize 1}    \rput(8,-0.5){\scriptsize 2}
\rput(10,-0.5){\scriptsize 3}   \rput(12,-0.5){\scriptsize 4}
\rput(14,-0.5){\scriptsize 5}   \rput(16,-0.5){\scriptsize 6}
\rput(18,-0.5){\scriptsize 7}   \rput(20,-0.5){\scriptsize 8}
\rput(22,-0.5){\scriptsize 9}   \rput(24,-0.5){\scriptsize 10}

\psline(0,2)(4,6)
\psline(4,6)(5,5)
\psline(5,5)(6,6)
\psline(6,6)(7,5)
\psline(7,5)(10,8)
\psline(10,8)(11,7)
\psline(11,7)(12,8)
\psline(12,8)(20,0)
\psline(20,0)(22,2)
\psline(22,2)(23,1)
\psline(23,1)(24,2)

}
\end{pspicture}
\end{center}
\end{figure}
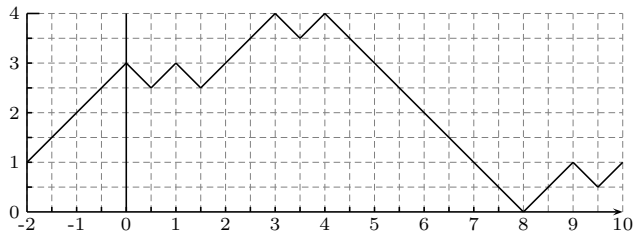

To illustrate this construction, consider the path
$\halfh\in\HH^5_{3,1}$ represented by the solid line in
Fig.~\ref{Fig:UnNorm}.
The minimal weight path $\halfh^{\mathrm{gs}}\in\HH^5_{3,1}$ is shown
dashed. 
Using \eqref{Eq:StraightWt} and \eqref{Eq:HalfWt}, we obtain
\begin{equation}
\begin{split}
\unwt(\halfh)&=\tfrac12(2+\tfrac52+\tfrac92+5+\tfrac{11}2+6
 +\tfrac{13}2+7+\tfrac{15}2+\tfrac{17}2)=\tfrac{55}2,\\
\unwt(\halfh^{\mathrm{gs}})&=\tfrac12(\tfrac12+1+\tfrac32+2)
=\tfrac72\quad \implies \quad
\wh(\halfh)=\tfrac{55}2-\tfrac72=24.
\end{split}
\end{equation}
Alternatively, we may use
the extended path shown in Fig.~\ref{Fig:ReNorm}.
From this, the renormalised weight $\wh(\halfh)$ is immediately
obtained via
\begin{equation}\wh(\halfh)=
 \tfrac12(-2-\tfrac32-1-\tfrac12
 +2+\tfrac52+\tfrac92+5+\tfrac{11}2+6
 +\tfrac{13}2+7+\tfrac{15}2+\tfrac{17}2)=24.\end{equation}

In \cite{PMnpd}, the generating function for
these paths was conjectured to be a Virasoro character:
\begin{equation}\label{Eq:GFconjecture}
\sum_{\halfh\in\HH^{p}_{\hat a,\hat b}} q^{\wh(\halfh)}
=\chi^{p,2p+1}_{r,s}(q),
\end{equation}
where the module labels $r,s$ are {given} by
\begin{equation}\label{rsabh}
s=2\hat a+1\quad \text{and}\quad r=\hat b.
\end{equation}

\subsection{Statement of the results}
In the following two sections, 
we describe a weight-preserving bijection between the sets
\begin{equation}\label{Eq:Bijection}
\PP^{p,2p+1}_{a,b}\leftrightarrow \HH^p_{\frac12(a-1),\frac12(b-1)} ,
\end{equation}
for $p\ge2$, and $a$ and $b$ odd
integers with $1\le a<2p$ and $3\le b<2p$.

In combination with the $p'=2p+1$ case of \eqref{Eq:GFvirasoro},
the establishment of this bijection proves \eqref{Eq:GFconjecture}.
Note that for the two sets related by \eqref{Eq:Bijection},
the values of the module labels $r$ and $s$,
obtained for the two cases using (\ref{rsab}) and (\ref{rsabh}) respectively,
are in agreement.

At first sight, the restriction on the parity of the values of $a$
and $b$ appears to constrain the applicability of our construction
to those modules which are labelled by $(r,s)$ with $s$ odd.
However, the equivalence of the modules labelled by $(r,s)$
and $(p-r,p'-s)$
(a consequence of the identity $h_{r,s}=h_{p-r,p'-s}$
for conformal dimensions)
implies that, in this $p'=2p+1$ case, the construction applies
to all inequivalent modules.

In what follows, it is notationally convenient to
make the identification
\begin{equation}
\PP^{p}_{a,b}\equiv \PP^{p,2p+1}_{a,b}.
\end{equation}

\section{The combinatorial bijection}
\label{Sec:CoBi}

The description of the bijection given in this section is
combinatorial because it
is specified in terms of direct manipulations of the paths.
This contrasts with the second description, presented in the next section,
which is formulated in terms of (non-local) operators.
Roughly, in the combinatorial scheme, a path from $\PP^p_{a,b}$
is transformed by first stripping off its (charge 1) particles,
then reinterpreting the cut path as a $\HH^p_{\hat a,\hat b}$
path through rescaling it by a factor of 1/2,
and finally, reinserting the particles in a precise way.

Throughout this section, we take $a$ and $b$ to be odd integers,
with $1\le a<2p$ and $3\le b<2p$,
and set 
\begin{equation}\label{Eq:ExtremeConvert}
\hat a=\frac12(a-1)\qquad\text{and}\qquad \hat b=\frac12(b-1).
\end{equation}

\subsection{Specifying the bijection}
\label{Sec:BijMap}

Let $h\in\PP^{p}_{a,b}$.
From $h$, repeatedly remove adjacent pairs of scoring vertices,
in each case adjoining the loose ends (which will be at the same
height), until no adjacent pair of scoring vertices remains.
Let $\hcut$ denote the resulting path.
For the path $h\in\PP^{4}_{1,3}$ given in
Fig.~\ref{TypicalScoringFig}, the resulting $\hcut\in\PP^{4}_{1,3}$
is given in Fig.~\ref{TypicalScoringCut}, having removed
$n=4$ pairs of scoring vertices from the former.

\begin{figure}[ht]
\caption{{\footnotesize $h^{\text{cut}}$ obtained
           from Fig. \ref{TypicalScoringFig}.}}
\vskip-1.5cm
\label{TypicalScoringCut}
\begin{center}
\begin{pspicture}(1,0)(13,5.5)

{
\psset{linestyle=none}
\psset{fillstyle=solid}
\psset{fillcolor=lightgray}
\psframe(0.5,1)(13.5, 1.5)
\psframe(0.5,2)(13.5, 2.5)
\psframe(0.5,3)(13.5, 3.5)
}

\psset{linestyle=solid}
\psline{-}(0.5,0.5)(13.5,0.5)
\psline{-}(0.5,4)(13.5,4)

\psline{-}(0.5,0.5)(0.5,4)
\psline{-}(13.5,0.5)(13.5,4)

{
\psset{linewidth=0.25pt,linestyle=dashed, dash=2.5pt 1.5pt,linecolor=gray}

\psline{-}(0.5,1.0)(13.5,1.0)
\psline{-}(0.5,1.5)(13.5,1.5)
\psline{-}(0.5,2.0)(13.5,2.0)
\psline{-}(0.5,2.5)(13.5,2.5)
\psline{-}(0.5,3)(13.5,3)
\psline{-}(0.5,3.5)(13.5,3.5)

\psline{-}(1.0,0.5)(1.0,4.0) \psline{-}(1.5,0.5)(1.5,4.0) 
\psline{-}(2.0,0.5)(2.0,4.0) \psline{-}(2.5,0.5)(2.5,4.0) \psline{-}(3.0,0.5)(3.0,4.0) 
\psline{-}(3.5,0.5)(3.5,4.0) \psline{-}(4.0,0.5)(4.0,4.0) \psline{-}(4.5,0.5)(4.5,4.0) 
\psline{-}(5.0,0.5)(5.0,4.0) \psline{-}(5.5,0.5)(5.5,4.0) \psline{-}(6.0,0.5)(6.0,4.0) 
\psline{-}(6.5,0.5)(6.5,4.0) \psline{-}(7.0,0.5)(7.0,4.0) \psline{-}(7.5,0.5)(7.5,4.0) 
\psline{-}(8.0,0.5)(8.0,4.0) \psline{-}(8.5,0.5)(8.5,4.0) \psline{-}(9.0,0.5)(9.0,4.0) 
\psline{-}(9.5,0.5)(9.5,4.0) \psline{-}(10.0,0.5)(10.0,4.0) \psline{-}(10.5,0.5)(10.5,4.0) 
\psline{-}(11.0,0.5)(11.0,4.0) \psline{-}(11.5,0.5)(11.5,4.0) \psline{-}(12.0,0.5)(12.0,4.0) 
\psline{-}(12.5,0.5)(12.5,4.0) \psline{-}(13.0,0.5)(13.0,4.0)  \psline{-}(13.5,0.5)(13.5,4.0)
}

{
\rput(0.25,0.5){{\scriptsize $1$}}
\rput(0.25,1.0){{\scriptsize $2$}} \rput(0.25,1.5){{\scriptsize $3$}}
\rput(0.25,2.0){{\scriptsize $4$}} \rput(0.25,2.5){{\scriptsize $5$}}
\rput(0.25,3.0){{\scriptsize $6$}} \rput(0.25,3.5){{\scriptsize $7$}}
\rput(0.25,4.0){{\scriptsize $8$}}

\rput(0.5,0.25){{\scriptsize $0$}}
\rput(1.0,0.25){{\scriptsize $1$}} \rput(1.5,0.25){{\scriptsize $2$}} \rput(2.0,0.25){{\scriptsize $3$}}
\rput(2.5,0.25){{\scriptsize $4$}} \rput(3.0,0.25){{\scriptsize $5$}} \rput(3.5,0.25){{\scriptsize $6$}}
\rput(4.0,0.25){{\scriptsize $7$}} \rput(4.5,0.25){{\scriptsize $8$}} \rput(5.0,0.25){{\scriptsize $9$}} 
\rput(5.5,0.25){{\scriptsize $10$}} \rput(6.0,0.25){{\scriptsize $11$}} \rput(6.5,0.25){{\scriptsize $12$}}
\rput(7.0,0.25){{\scriptsize $13$}} \rput(7.5,0.25){{\scriptsize $14$}} \rput(8.0,0.25){{\scriptsize $15$}}
\rput(8.5,0.25){{\scriptsize $16$}} \rput(9.0,0.25){{\scriptsize $17$}} \rput(9.5,0.25){{\scriptsize $18$}}
\rput(10.0,0.25){{\scriptsize $19$}} \rput(10.5,0.25){{\scriptsize $20$}} \rput(11.0,0.25){{\scriptsize $21$}}   
\rput(11.5,0.25){{\scriptsize $22$}} \rput(12.0,0.25){{\scriptsize $23$}} \rput(12.5,0.25){{\scriptsize $24$}}
\rput(13.0,0.25){{\scriptsize $25$}} \rput(13.5,0.25){{\scriptsize $26$}} 
}

{
\psset{linestyle=solid}

\psline{-}(.5,.5)(3.5,3.5) 
\psline{-}(3.5,3.5)(4.0,3.0) \psline{-}(4.0,3.0)(4.5,3.5)
\psline{-}(4.5,3.5)(6.0,2.0) \psline{-}(6.0,2.0)(6.5,2.5)
\psline{-}(6.5,2.5)(8.0,1)
\psline{-}(8.0,1)(8.5,1.5) \psline{-}(8.5,1.5)(9.0,1.0)
\psline{-}(9.0,1)(9.5,1.5) \psline{-}(9.5,1.5)(10.0,1.0)
\psline{-}(10.0,1)(10.5,1.5) \psline{-}(10.5,1.5)(11.0,1.0)
\psline{-}(11.0,1)(11.5,1.5) \psline{-}(11.5,1.5)(12.0,1.0)
\psline{-}(12.0,1)(12.5,1.5) \psline{-}(12.5,1.5)(13.0,1.0)
\psline{-}(13.0,1)(13.5,1.5) 
}
{
\psset{dotsize=3.5pt}
\psdots(5.5,2.5)(7.5,1.5)
\psset{fillcolor=white}
\psset{dotsize=4pt}\psset{dotstyle=o}
\psdots(1,1)(2,2)(3,3)
}
\end{pspicture}
\end{center}
\end{figure}

To describe the bijection, it is also required that, when carrying
out the above removal process, we record the number of non-scoring
vertices to the left of each pair of scoring vertices that is removed.
This resulting list of $n$ integers then encodes a partition,
of at most $n$ parts, which we denote $\lambda(h)$.
In the case of the path $h$ of Fig.~\ref{TypicalScoringFig},
the $n=4$ removals result in the partition $\lambda(h)=(9,8,5,1)$.

Note that the removal process ensures that $\hcut$ has
no peak at an even height, and no valley at an odd height.
Therefore, if we shrink the whole path $\hcut$ by a factor of 2,
and decrease the values of the heights by 1/2, we obtain a
half-lattice path, $\hlac$, whose peaks are all at integer heights
and whose valleys are all at non-integer heights.
In particular, $\hlac$ does not extend above height $p-1$.
In addition, $\hlac_0=\hat a$ and $\hlac$ is $\hat b$-tailed.
Therefore, $\hlac\in\HH^p_{\hat a,\hat b}$.
In the case of the $\hcut$ of Fig.~\ref{TypicalScoringCut},
the resulting $\hlac$ is given in Fig.~\ref{TypicalScoringHat}.

\begin{figure}[ht]
\caption{{\footnotesize $\hh^{\text{cut}}$ obtained from
  Fig. \ref{TypicalScoringCut}}}
\label{TypicalScoringHat}
\begin{center}
\begin{pspicture}(2.3,-0.5)(12.5,2.3)
{\psset{yunit=0.32cm,xunit=0.32cm,linewidth=.6pt}

{
\psset{linewidth=0.25pt,linestyle=dashed, dash=2.5pt 1.5pt,linecolor=gray}

\psline{-}(0,1)(46,1.0)
\psline{-}(0,2)(46,2.0)
\psline{-}(0,3)(46,3.0)
\psline{-}(0,4)(46,4.0)
\psline{-}(0,5)(46,5.0)
\psline{-}(0,6)(46,6.0)

\psline{-}(1,0)(1,6) \psline{-}(2,0)(2,6) \psline{-}(3,0)(3,6) 
\psline{-}(4,0)(4,6) \psline{-}(5,0)(5,6) \psline{-}(6,0)(6,6)
\psline{-}(7,0)(7,6) \psline{-}(8,0)(8,6) \psline{-}(9,0)(9,6) 
\psline{-}(10,0)(10,6) \psline{-}(11,0)(11,6) \psline{-}(12,0)(12,6)
\psline{-}(13,0)(13,6) \psline{-}(14,0)(14,6) \psline{-}(15,0)(15,6) 
\psline{-}(16,0)(16,6) \psline{-}(17,0)(17,6) \psline{-}(18,0)(18,6)
\psline{-}(19,0)(19,6) \psline{-}(20,0)(20,6) \psline{-}(21,0)(21,6) 
\psline{-}(22,0)(22,6) \psline{-}(23,0)(23,6) \psline{-}(24,0)(24,6) \psline{-}(25,0)(25,6)
\psline{-}(26,0)(26,6) \psline{-}(27,0)(27,6) \psline{-}(28,0)(28,6) 
\psline{-}(29,0)(29,6) \psline{-}(30,0)(30,6) \psline{-}(31,0)(31,6)
\psline{-}(32,0)(32,6) \psline{-}(33,0)(33,6) \psline{-}(34,0)(34,6) 
\psline{-}(35,0)(35,6) \psline{-}(36,0)(36,6) \psline{-}(37,0)(37,6)
\psline{-}(38,0)(38,6) \psline{-}(39,0)(39,6) \psline{-}(40,0)(40,6) 
\psline{-}(41,0)(41,6) \psline{-}(42,0)(42,6) \psline{-}(43,0)(43,6)
\psline{-}(44,0)(44,6) \psline{-}(45,0)(45,6) \psline{-}(46,0)(46,6)

}

{
\psline{-}(0,0)(0,6)
\psline{->}(0,0)(46,0)

\psline(0,1)(0.2,1)
\psline(0,2)(0.2,2)
\psline(0,3)(0.2,3)
\psline(0,4)(0.2,4)
\psline(0,5)(0.2,5)
\psline(0,6)(0.5,6)

\psline(1,0)(1,.2)     \psline(2,0)(2,.2)        \psline(3,0)(3,.2)
\psline(4,0)(4,.2)     \psline(5,0)(5,.2)        \psline(6,0)(6,.2)
\psline(7,0)(7,.2)     \psline(8,0)(8,.2)        \psline(9,0)(9,.2)
\psline(10,0)(10,.2)     \psline(11,0)(11,.2)        \psline(12,0)(12,.2)
\psline(13,0)(13,.2)     \psline(14,0)(14,.2)        \psline(15,0)(15,.2)
\psline(16,0)(16,.2)     \psline(17,0)(17,.2)        \psline(18,0)(18,.2)
\psline(19,0)(19,.2)     \psline(20,0)(20,.2)        \psline(21,0)(21,.2)
\psline(22,0)(22,.2)    \psline(23,0)(23,.2)    \psline(24,0)(24,.2)    
\psline(25,0)(25,.2)   \psline(26,0)(26,.2)    \psline(27,0)(27,.2)   
\psline(28,0)(28,.2)   \psline(29,0)(29,.2)    \psline(30,0)(30,.2)
\psline(31,0)(31,.2)   \psline(32,0)(32,.2)    \psline(33,0)(33,.2)
\psline(34,0)(34,.2)   \psline(35,0)(35,.2)    \psline(36,0)(36,.2)
\psline(37,0)(37,.2)   \psline(38,0)(38,.2)    \psline(39,0)(39,.2)
\psline(40,0)(40,.2)   \psline(41,0)(41,.2)    \psline(42,0)(42,.2)
\psline(43,0)(43,.2)   \psline(44,0)(44,.2)    \psline(45,0)(45,.2)
}

{
\rput(-.5,0){\scriptsize 0}
\rput(-.5,2){\scriptsize 1}
\rput(-.5,4){\scriptsize 2}
\rput(-.5,6){\scriptsize 3}

\rput(0,-0.5){\scriptsize 0}
\rput(2,-0.5){\scriptsize 1}    \rput(4,-0.5){\scriptsize 2}
\rput(6,-0.5){\scriptsize 3}    \rput(8,-0.5){\scriptsize 4}
\rput(10,-0.5){\scriptsize 5}   \rput(12,-0.5){\scriptsize 6}
\rput(14,-0.5){\scriptsize 7}   \rput(16,-0.5){\scriptsize 8}
\rput(18,-0.5){\scriptsize 9}   \rput(20,-0.5){\scriptsize 10} 
\rput(22,-0.5){\scriptsize 11}  \rput(24,-0.5){\scriptsize 12}
\rput(26,-0.5){\scriptsize 13}  \rput(28,-0.5){\scriptsize 14}
\rput(30,-0.5){\scriptsize 15}  \rput(32,-0.5){\scriptsize 16}
\rput(34,-0.5){\scriptsize 17}  \rput(36,-0.5){\scriptsize 18}
\rput(38,-0.5){\scriptsize 19}  \rput(40,-0.5){\scriptsize 20}
\rput(42,-0.5){\scriptsize 21}  \rput(44,-0.5){\scriptsize 22}
}

{
\psline(0,0)(6,6)
\psline(6,6)(7,5)
\psline(7,5)(8,6)
\psline(8,6)(11,3)
\psline(11,3)(12,4)
\psline(12,4)(15,1)
\psline(15,1)(16,2)
\psline(16,2)(17,1)
\psline(17,1)(18,2)
\psline(18,2)(19,1)
\psline(19,1)(20,2)
\psline(20,2)(21,1)
\psline(21,1)(22,2)
\psline(22,2)(23,1)
\psline(23,1)(24,2)
\psline(24,2)(25,1)
\psline(25,1)(26,2)
\psline(26,2)(27,1)
\psline(27,1)(28,2)
\psline(28,2)(29,1)
\psline(29,1)(30,2)
\psline(30,2)(31,1)
\psline(31,1)(32,2)
\psline(32,2)(33,1)
\psline(33,1)(34,2)
\psline(34,2)(35,1)
\psline(35,1)(36,2)
\psline(36,2)(37,1)
\psline(37,1)(38,2)
\psline(38,2)(39,1)
\psline(39,1)(40,2)
\psline(40,2)(41,1)
\psline(41,1)(42,2)
\psline(42,2)(43,1)
\psline(43,1)(44,2)
\psline(44,2)(45,1)
\psline(45,1)(46,2)
}
{

}
}
\end{pspicture}
\end{center}
\end{figure}

As already stressed, each valley of $\hlac$ occurs at a
non-integer height.
The path $\hat h$ is obtained from $\hat h^{\mathrm{cut}}$ by
deepening
$n$ of these valleys to the next integer height
by inserting pairs of SE-NE edges.
On labelling the valleys of $\hlac$ from
left to right by $1, 2, 3,\ldots,$ the $n$ valleys labelled
$\mu_1$, $\mu_{2},\ldots,\mu_n$,
are deepened, where we set
\begin{equation}\label{Eq:StaggerParts}
\mu_i=\lambda_i+n+1-i,
\end{equation}
with $\lambda=\lambda(h)$.
Note that the partition $\mu=(\mu_1,\mu_2,\ldots,\mu_n)$
has distinct parts.

In our ongoing example, we have $\lambda=(9,8,5,1)$.
{}From this \eqref{Eq:StaggerParts} yields $\mu=(13,11,7,2)$,
and therefore
we obtain $\hat h$ from $\hlac$ by deepening the
$13^{\text{th}}$, $11^{\text{th}}$, $7^{\text{th}}$ and $2^{\text{nd}}$
valleys of the latter.
The resulting path $\hat h$ is given in Fig.~\ref{che_h}.
It may be checked that $\wh(\hat h)=74=\wt(h)$ in this case.

We claim that the combined map
$h\to(\hcut,n,\lambda)\to(\hlac,n,\mu)\to\hat h$
is a weight-preserving bijection between
$\PP^{p}_{a,b}$ and $\HH^p_{\hat a,\hat b}$.
That it is a bijection follows because the inverse map from 
$\HH^p_{\hat a,\hat b}$ to $\PP^{p}_{a,b}$,
which is easily described, is well-defined.
This relies on the fact that each $\hat h\in\HH^p_{\hat a,\hat b}$
arises from a unique path, $\hlac$, whose characteristic
property, we recall, is that its valleys are all at
non-integer heights;
$\hlac$ is thus recovered from $\hat h$ by making shallow each
integer valley.
With $n$ the number of such integer valleys,
the partition $\mu=(\mu_1,\ldots,\mu_n)$ is determined by
setting its parts to be the numberings of the integer valleys
amongst all valleys, counted from the left.
The parts of $\mu$ are necessarily distinct, and thus
a genuine partition $\lambda$ is recovered via
\eqref{Eq:StaggerParts}.
That the map $h\to\hat h$ is weight-preserving is demonstrated below.

\subsection{Removing basic particles from the RSOS paths%
\protect\footnote{
  The process presented in this section is described,
  using similar terminology, in \cite[Section 5]{TWpa}
  (cf.\ eqn.~(15) therein), and previously in 
  \cite[Section 2]{FLPW} using the notions of $\BB_2$ and $\BB_3$ transforms
  (therein, $h^{(0)}$ is used to denote $h^{\mathrm{cut}}$).}}
\label{Sec:RemFB}

Each pair of adjacent scoring vertices
(which are necessarily of different types)
in $h\in\PP^{p}_{a,b}$ is identified with a particle:%
\footnote{These are the particles of charge 1 in the
           terminology of \cite{PMprsos}.}
where there occur $d\ge2$ consecutive scoring vertices,
we identify $\lfloor d/2\rfloor$ particles
(when $d$ is odd, the ambiguity over which pairs are the actual
particles is immaterial).
The excitation of each of these particles is defined to be
the number of non-scoring vertices to its left.
Thus, the partition $\lambda(h)$, defined in Section \ref{Sec:BijMap},
lists the excitations of the particles in $h$.


A path which contains no particles is said to be
\emph{particle-deficient}.
We construct the particle-deficient path
$h^{\mathrm{cut}}\in\PP^{p}_{a,b}$ simply by removing all of the
particles from $h\in\PP^{p}_{a,b}$.
To determine the weight of $h^{\mathrm{cut}}$, first
consider the removal of one such particle from $h$.
Let $\lambda_i$ be the number of non-scoring vertices to its left,
and let $k$ be the total number of scoring vertices in $h$ 
($k=13$ for the path of Fig.~\ref{TypicalScoringFig}).
The band structure for the $p'=2p+1$ cases
ensures that the first of the two scoring vertices that
comprise the particle is necessarily a peak or valley.
Let it be at position $(x,h_x)$.
If it is a peak, then the following vertex is at $(x+1,h_x-1)$,
and together they contribute
\begin{equation}
u_x+v_{x+1}=\frac12(x-h_x+a+x+1+h_x-1-a)=x
\end{equation} to the weight.
If it is a valley, then the following vertex is at $(x+1,h_x+1)$,
and together they contribute
\begin{equation}
v_x+u_{x+1}=\frac12(x+h_x-a+x+1-h_x-1+a)=x
\end{equation}
 to the weight.
There are $x-1-\lambda_i$ scoring vertices to the left of
the particle, and thus $\lambda_i+k-x-1$ to its right.
On removing the particle, the contribution of each of the latter 
to the weight decreases by one.
Thus the total weight reduction on removing the particle is
$\lambda_i+k-1$.

Then, if $h$ contains $n$ particles, on removing all of them,
noting that $k$ decreases by
two at each step, we obtain
\begin{equation}\label{Eq:RemFB}
\begin{split}
\wt(h^{\mathrm{cut}})
&=\wt(h) - \sum_{i=1}^n \lambda_i - (k-1)-(k-3)-(k-5)-\cdots-(k-2n-1)\\
&=\wt(h) - \sum_{i=1}^n \lambda_i - n(k-n).
\end{split}
\end{equation}


\subsection{Mapping from RSOS paths to half-lattice paths}
\label{Sec:CutMap}
${}$
Given a particle-deficient path $\hcut\in\PP^{p}_{a,b}$, define a
half-lattice path $\hlac$ by shrinking $\hcut$
by a factor of 2, and decreasing all heights by 1/2.
Note that
$\hlac$ is a element of $\HH^p_{\hat a,\hat b}$,
whose valleys are at non-integer heights.
In this section, we show that $\wh(\hlac)=\wt(\hcut)$.

\begin{figure}[ht]
\caption{{\footnotesize Example of segment matching for a
path {$h$}. 
}}
\vskip-1.2cm
\label{TypicalMatching}
\begin{center}
\begin{pspicture}(1,0)(11,6.5)

{
\psset{linestyle=none}
\psset{fillstyle=solid}
\psset{fillcolor=lightgray}
\psframe(0.5,1)(11.5, 1.5)
\psframe(0.5,2)(11.5, 2.5)
\psframe(0.5,3)(11.5, 3.5)
\psframe(0.5,4)(11.5, 4.5)
}

\psset{linestyle=solid}
\psline{-}(0.5,0.5)(11.5,0.5)
\psline{-}(0.5,5)(11.5,5)

\psline{-}(0.5,0.5)(0.5,5)
\psline{-}(11.5,0.5)(11.5,5)

{
\psset{linewidth=0.25pt,linestyle=dashed, dash=2.5pt 1.5pt,linecolor=gray}

\psline{-}(0.5,1.0)(11.5,1.0)
\psline{-}(0.5,1.5)(11.5,1.5)
\psline{-}(0.5,2.0)(11.5,2.0)
\psline{-}(0.5,2.5)(11.5,2.5)
\psline{-}(0.5,3)(11.5,3)
\psline{-}(0.5,3.5)(11.5,3.5)
\psline{-}(0.5,4)(11.5,4)
\psline{-}(0.5,4.5)(11.5,4.5)

\psline{-}(1.0,0.5)(1.0,5.0) \psline{-}(1.5,0.5)(1.5,5.0) 
\psline{-}(2.0,0.5)(2.0,5.0) \psline{-}(2.5,0.5)(2.5,5.0) \psline{-}(3.0,0.5)(3.0,5.0) 
\psline{-}(3.5,0.5)(3.5,5.0) \psline{-}(4.0,0.5)(4.0,5.0) \psline{-}(4.5,0.5)(4.5,5.0) 
\psline{-}(5.0,0.5)(5.0,5.0) \psline{-}(5.5,0.5)(5.5,5.0) \psline{-}(6.0,0.5)(6.0,5.0) 
\psline{-}(6.5,0.5)(6.5,5.0) \psline{-}(7.0,0.5)(7.0,5.0) \psline{-}(7.5,0.5)(7.5,5.0) 
\psline{-}(8.0,0.5)(8.0,5.0) \psline{-}(8.5,0.5)(8.5,5.0) \psline{-}(9.0,0.5)(9.0,5.0) 
\psline{-}(9.5,0.5)(9.5,5.0) \psline{-}(10.0,0.5)(10.0,5.0) \psline{-}(10.5,0.5)(10.5,5.0) 
\psline{-}(11.0,0.5)(11.0,5.0) \psline{-}(11.5,0.5)(11.5,5.0) 
}

{
\rput(0.25,0.5){{\scriptsize $1$}}
\rput(0.25,1.0){{\scriptsize $2$}} \rput(0.25,1.5){{\scriptsize $3$}}
\rput(0.25,2.0){{\scriptsize $4$}} \rput(0.25,2.5){{\scriptsize $5$}}
\rput(0.25,3.0){{\scriptsize $6$}} \rput(0.25,3.5){{\scriptsize $7$}}
\rput(0.25,4.0){{\scriptsize $8$}} \rput(0.25,4.5){{\scriptsize $9$}}
\rput(0.25,5.0){{\scriptsize $10$}}

\rput(0.5,0.25){{\scriptsize $0$}}
\rput(1.0,0.25){{\scriptsize $1$}} \rput(1.5,0.25){{\scriptsize $2$}} \rput(2.0,0.25){{\scriptsize $3$}}
\rput(2.5,0.25){{\scriptsize $4$}} \rput(3.0,0.25){{\scriptsize $5$}} \rput(3.5,0.25){{\scriptsize $6$}}
\rput(4.0,0.25){{\scriptsize $7$}} \rput(4.5,0.25){{\scriptsize $8$}} \rput(5.0,0.25){{\scriptsize $9$}} 
\rput(5.5,0.25){{\scriptsize $10$}} \rput(6.0,0.25){{\scriptsize $11$}} \rput(6.5,0.25){{\scriptsize $12$}}
\rput(7.0,0.25){{\scriptsize $13$}} \rput(7.5,0.25){{\scriptsize $14$}} \rput(8.0,0.25){{\scriptsize $15$}}
\rput(8.5,0.25){{\scriptsize $16$}} \rput(9.0,0.25){{\scriptsize $17$}} \rput(9.5,0.25){{\scriptsize $18$}}
\rput(10.0,0.25){{\scriptsize $19$}} \rput(10.5,0.25){{\scriptsize $20$}} \rput(11.0,0.25){{\scriptsize $21$}}   
\rput(11.5,0.25){{\scriptsize $22$}} 
}

{
\psset{linestyle=solid}

\psline{-}(.5,2.5)(2.5,4.5) 
\psline{-}(2.5,4.5)(3.0,4.0) \psline{-}(3.0,4.0)(3.5,4.5)
\psline{-}(3.5,4.5)(5,3) \psline{-}(5,3)(6.5,4.5)
\psline{-}(6.5,4.5)(10,1) \psline{-}(10,1)(11.5,2.5)
}

{
\psset{linewidth=0.5pt,linestyle=solid,linecolor=black}
\psline{<->}(0.9,2.75)(8.1,2.75)
\psline{<->}(1.4,3.25)(4.6,3.25)  \psline{<->}(5.4,3.25)(7.6,3.25)
\psline{<->}(1.9,3.75)(4.1,3.75)  \psline{<->}(5.9,3.75)(7.1,3.75)
\psline{<->}(2.35,4.25)(2.65,4.25)  \psline{<->}(3.35,4.25)(3.65,4.25)  \psline{<->}(6.35,4.25)(6.65,4.25)
\psline{<->}(8.9,2.25)(11.1,2.25)
\psline{<->}(9.4,1.75)(10.6,1.75)
\psline{<->}(9.85,1.25)(10.15,1.25)

}

\end{pspicture}
\end{center}
\end{figure}
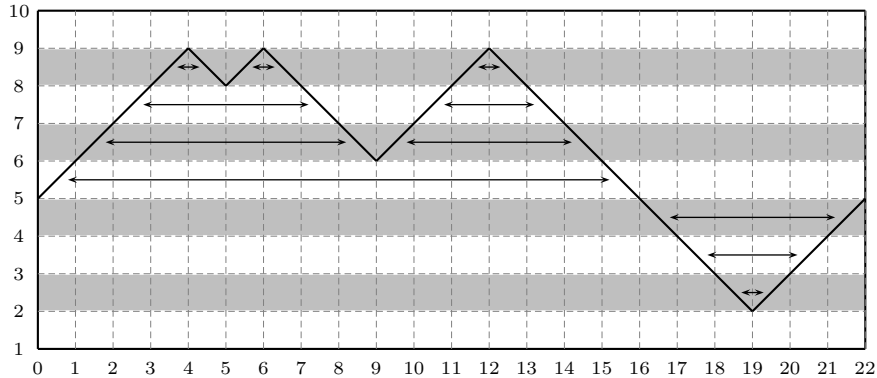

First consider the case where $a=b$.
(The path $\hcut$ then starts and ends at the same height
--- if truncated beyond the start of the final oscillation).
In such a case, we may match (pair) each path segment with another
at the same height, one NE and one SE (the order is immaterial).
This matching process is illustrated in Fig.~\ref{TypicalMatching}.


Since $\hcut$ has no peak at an even height and no valley
at an odd height, the scoring vertices of $\hcut$ occur
at the right ends of all the segments in the light bands.
Consider a specific matched pair of segments
in a light band,
with the left ends of the
NE and SE segments at positions $(x,y)$ and $(x',y+1)$ respectively.
The two scoring vertices at $(x+1,y+1)$ and $(x'+1,y)$ together contribute
\begin{equation}
u_{x+1}+v_{x'+1}=\frac12(x+1-(y+1)+a+x'+1+y-a)=\frac12(x+x'+1)
\end{equation}
to $\wt(\hcut)$.

For the corresponding half-lattice path $\hlac$,
the weight may be obtained by considering the four straight
vertices at each end of each of the segments of the matched pair
(the vertex at $(0,\hat a)$ will be required here if it's straight).
These four vertices contribute
\begin{equation}
\frac12\left[\frac12x+\frac12(x+1)+\frac12x'+\frac12(x'+1)\right]=\frac12(x+x'+1)
\end{equation}
to $\unwt(\hlac)$.
Since this agrees with the contribution of the corresponding
two scoring vertices of $\hcut$ to the weight
$\wt(\hcut)$, we conclude that $\unwt(\hlac)=\wt(\hcut)$.
But, for $a=b$, $\wh(\hlac)=\unwt(\hlac)$,
thereby proving 
that $\wh(\hlac)=\wt(\hcut)$
in this $a=b$ case.

In the case that $a\ne b$, we make use of the trick described in
Section \ref{Sec:Halfpath} to obtain $\wh(\hlac)$ by extending the
path $\hlac$ to the left.
On the other hand,
extending the path $\hcut$ to the left by $2e=\vert a-b\vert$ steps
with $\hcut_{-2e}=b$, creates a path of unchanged weight $\wt(\hcut)$
because, via \eqref{Eq:ScoreWts}, the additional scoring vertices
each contribute 0 to the weight.
Then, upon applying the argument used above in the $a=b$ case to
these extended paths, we obtain $\wh(\hlac)=\wt(\hcut)$ for
$a\ne b$ also.

\subsection{Deepening valleys}
\label{Sec:DeeperVal}
Consider a half-lattice path
$\hat h^{(0)}\in\mathcal H^p_{\hat a,\hat b}$,
having $m^{}$ straight vertices. 
This count includes consideration of the vertex at $(0,\hat a)$,
which, through the convention stated in Section \ref{Sec:Halfpath},
is deemed straight if and only if the first segment of $\hat h^{(0)}$
is in the NE direction.
The vertices of $\hat h^{(0)}$ that do not contribute to
its weight are the peaks and valleys.
We now determine the change in weight on deepening one of the valleys.
Let the valley being deepened be the $j$th, counting from the left,
and let it be situated at $(x,\hat h^{(0)}_x)$.
There are necessarily $j$ peaks to the left of this valley
(perhaps including one at $(0,\hat a)$), and therefore
$2x+1-2j$ straight vertices.
After this position, there are thus $m^{}-2x+2j-1$ straight vertices.
The deepening moves each of these to the right by two
(half-integer) positions.
It also introduces two straight vertices, at positions
$(x,\hat h^{(0)}_x)$ and $(x+1,\hat h^{(0)}_x)$.
Thus, if this resulting path is denoted $\hat h^{(1)}$,
\begin{equation}
\begin{split}
\wh(\hat h^{(1)})
&=\wh(\hat h^{(0)})
+\frac{1}{2} (x+(x+1))+\frac12(m^{}-2x+2j-1))\\
&=\wh(\hat h^{(0)}) +\frac{1}{2}m^{} +j.
\end{split}
\end{equation}

Note that the deepening increases the value of $m^{}$ by 2.
So if we obtain the path $\hat h$ by performing a succession of
deepenings to a path $\hlac$ at valleys numbered
$\mu_1$, $\mu_2$, $\ldots,\mu_n$,
we have
\begin{equation}\label{Eq:DeeperVal}
\begin{split}
\wh(\hat h)
&=\wh(\hlac)
+\frac{1}{2} (m^{}+(m^{}+2)+\cdots+(m^{}+2n-2))
+\sum_{i=1}^n \mu_i\\
&=\wh(\hlac)
+\frac{n}{2} (m^{}+n-1) + \sum_{i=1}^n \mu_i.
\end{split}
\end{equation}

\subsection{Altogether now}

Consider the combined map
$h\to(\hcut,n,\lambda)\to(\hlac,n,\mu)\to\hat h$
defined in Section \ref{Sec:BijMap} above.
Let $k$ and $k'$ be the number of
scoring vertices in $h$ and $\hcut$ respectively,
and let $m^{}$ be the number of straight vertices
(including consideration of the vertex at $(0,\hat a)$) in $\hlac$.
The pair removal process in Section \ref{Sec:RemFB} shows
that $k'=k-2n$.
The matching of edges described in Section \ref{Sec:CutMap}
shows that $m^{}=2k'$ and thus $m^{}=2k-4n$.
Then, using eqns.~\eqref{Eq:RemFB} and \eqref{Eq:DeeperVal}, and
the fact that $\wh(\hlac)=\wt(\hcut)$, we obtain
\begin{equation}\label{Eq:WtDiff}
\begin{split}
\wh(\hat h)-\wt(h)
&=\sum_{i=1}^n \mu_i -\sum_{i=1}^n \lambda_i
+\frac n2(m^{}+n-1)-n(k-n)\\
&=\sum_{i=1}^n \mu_i -\sum_{i=1}^n \lambda_i
-\frac n2(n+1)\\
&=0,
\end{split}
\end{equation}
where the final equality follows because,
using \eqref{Eq:StaggerParts},
\begin{equation}
\sum_{i=1}^n\mu_i-\sum_{i=1}^n\lambda_i
=\sum_{i=1}^n i=\frac n2(n+1).
\end{equation}
Thus the weight-preserving nature of the bijection has been proved.


\section{The path-operator bijection}
\label{Sec:OperatorBij}

In this section, we introduce natural descriptions of
each of the two types of paths
in terms of sequences of (local and non-local) operators.
In the case of the RSOS paths $\PP^{p}_{a,b}$, these sequences
are encodings of the vertex words introduced in \cite{TWpa}.
For the $\HH^p_{\hat a,\hat b}$ paths, the sequences are
simplified versions of a modification of those
already presented in \cite{PMnlob}.
The bijection between $\PP^{p}_{a,b}$ and $\HH^p_{\hat a,\hat b}$,
described in the previous section, is then formulated as a rule
for transforming between such sequences.


\subsection{$\PP^p$ paths as sequences of operators}
\label{Sec:RSOSop}

For each path $h\in\PP^{p,p'}_{a,b}$, the vertex word $\vword{h}$ is
defined to be the sequence of letters $S$ and $N$ which indicate
the sequence of vertices, scoring or non-scoring, of $h$,
read from the left \cite[\S5]{TWpa}.
For example, for the path $h$ given in Fig.~\ref{TypicalScoringFig},
\begin{equation}\label{Eq:TypicalVword}
\vword{h}=SNSSSNSNNNSSNSNNSSNSSSNNNNNNNNNNN\cdots.
\end{equation}
Each vertex word $\vword{h}$ is of infinite length and, because the
tail of $h$ lies in a dark band, $\vword{h}$ contains only a finite
number of letters $S$.
As indicated in \cite{TWpa}, every infinite length word $\vsymbol$
in $S$ and $N$ having only finite number of entries $S$,
is the vertex word $\vsymbol=\vword{h}$ of at most one path $h$.

\begin{figure}[ht]
\caption{{\footnotesize The action of the operators $\db_x$, $\cb_x$
and $\cbs_x$ on a portion of a path that oscillates in a dark band.
Note that $\db_x$ either creates a valley (cf.~(a)) or a peak (cf.~(b))
in a light band according to the odd/even parity of $x$.
}}
\vskip-1.4cm
\label{Fig:CCDaction}
\begin{center}

\begin{pspicture}(1,0)(9,3.5)  

{
\psset{linestyle=none}
\psset{fillstyle=solid}
\psset{fillcolor=lightgray}
\psframe(0.5,1)(3.5, 1.5)
\psframe(6.5,1)(9.5, 1.5)
}

{
\psset{linestyle=solid}
\psline{-}(0.5,0.5)(3.5,0.5)
\psline{-}(0.5,2)(3.5,2)
\psline{-}(0.5,0.5)(0.5,2)
\psline{-}(3.5,0.5)(3.5,2)

\psline{-}(6.5,0.5)(9.5,0.5)
\psline{-}(6.5,2)(9.5,2)
\psline{-}(6.5,0.5)(6.5,2)
\psline{-}(9.5,0.5)(9.5,2)

\psline{->}(4.5,1.25)(5.5,1.25) 
\rput(5.0,1.5){{\scriptsize $\db _{x}$}}
}

{
\psset{linewidth=0.25pt,linestyle=dashed, dash=2.5pt 1.5pt,linecolor=gray}

\psline{-}(0.5,1.0)(3.5,1.0)  \psline{-}(6.5,1.0)(9.5,1.0)
\psline{-}(0.5,1.5)(3.5,1.5)  \psline{-}(6.5,1.5)(9.5,1.5)

\psline{-}(1.0,0.5)(1.0,2.0) \psline{-}(1.5,0.5)(1.5,2.0) 
\psline{-}(2.0,0.5)(2.0,2.0) \psline{-}(2.5,0.5)(2.5,2.0) 
\psline{-}(3.0,0.5)(3.0,2.0) \psline{-}(3.5,0.5)(3.5,2.0) 

\psline{-}(7.0,0.5)(7.0,2.0) \psline{-}(7.5,0.5)(7.5,2.0) 
\psline{-}(8.0,0.5)(8.0,2.0) \psline{-}(8.5,0.5)(8.5,2.0)
\psline{-}(9.0,0.5)(9.0,2.0) \psline{-}(9.5,0.5)(9.5,2.0)
}


\rput(0.25,1.0){{\scriptsize $l$}}
\rput(0.1,1.5){{\scriptsize $l+1$}}
\rput(-0.5,2){{\scriptsize ${\bf{a)}}$}}
\rput(9.75,1.0){{\scriptsize $l$}}
\rput(10,1.5){{\scriptsize $l+1$}}

\rput(1.5,0.25){{\scriptsize $x$}}
\rput(7.5,0.25){{\scriptsize $x$}}

{
\psset{linestyle=solid}

\psline{-}(.5,1)(1.0,1.5) \psline{-}(1.0,1.5)(1.5,1)
\psline{-}(1.5,1)(2.0,1.5) \psline{-}(2.0,1.5)(2.5,1)
\psline{-}(2.5,1)(3.0,1.5) \psline{-}(3.0,1.5)(3.5,1)

\psline{-}(6.5,1)(7.0,1.5) \psline{-}(7.0,1.5)(7.5,1.0)
\psline{-}(7.5,1.0)(8.0,0.5) \psline{-}(8.0,0.5)(8.5,1)
\psline{-}(8.5,1)(9.0,1.5) \psline{-}(9.0,1.5)(9.5,1)
}
{
\psset{linewidth=0.25pt,linestyle=dashed, dash=2.5pt 1.5pt,linecolor=black}
\psline{-}(7.5,1)(8,1.5) \psline{-}(8,1.5)(8.5,1)
}
{
\psset{dotsize=3.5pt}
\psdots(8,0.5)
\psset{fillcolor=white}
\psset{dotsize=4pt}\psset{dotstyle=o}
\psdots(8.5,1)
}

\end{pspicture} 

\begin{pspicture}(1,0)(9,2.5) 

{
\psset{linestyle=none}
\psset{fillstyle=solid}
\psset{fillcolor=lightgray}
\psframe(0.5,1)(3.5, 1.5)
\psframe(6.5,1)(9.5, 1.5)
}

{
\psset{linestyle=solid}
\psline{-}(0.5,0.5)(3.5,0.5)
\psline{-}(0.5,2)(3.5,2)
\psline{-}(0.5,0.5)(0.5,2)
\psline{-}(3.5,0.5)(3.5,2)

\psline{-}(6.5,0.5)(9.5,0.5)
\psline{-}(6.5,2)(9.5,2)
\psline{-}(6.5,0.5)(6.5,2)
\psline{-}(9.5,0.5)(9.5,2)

\psline{->}(4.5,1.25)(5.5,1.25) 
\rput(5.0,1.5){{\scriptsize $\db _{x}$}}
}

{
\psset{linewidth=0.25pt,linestyle=dashed, dash=2.5pt 1.5pt,linecolor=gray}

\psline{-}(0.5,1.0)(3.5,1.0)  \psline{-}(6.5,1.0)(9.5,1.0)
\psline{-}(0.5,1.5)(3.5,1.5)  \psline{-}(6.5,1.5)(9.5,1.5)

\psline{-}(1.0,0.5)(1.0,2.0) \psline{-}(1.5,0.5)(1.5,2.0) 
\psline{-}(2.0,0.5)(2.0,2.0) \psline{-}(2.5,0.5)(2.5,2.0) 
\psline{-}(3.0,0.5)(3.0,2.0) \psline{-}(3.5,0.5)(3.5,2.0) 

\psline{-}(7.0,0.5)(7.0,2.0) \psline{-}(7.5,0.5)(7.5,2.0) 
\psline{-}(8.0,0.5)(8.0,2.0) \psline{-}(8.5,0.5)(8.5,2.0)
\psline{-}(9.0,0.5)(9.0,2.0) \psline{-}(9.5,0.5)(9.5,2.0)
}


\rput(0.25,1.0){{\scriptsize $l$}}
\rput(0.1,1.5){{\scriptsize $l+1$}}
\rput(-0.5,2){{\scriptsize ${\bf{b)}}$}}
\rput(9.75,1.0){{\scriptsize $l$}}
\rput(10,1.5){{\scriptsize $l+1$}}

\rput(1.5,0.25){{\scriptsize $x$}}
\rput(7.5,0.25){{\scriptsize $x$}}

{
\psset{linestyle=solid}

\psline{-}(.5,1.5)(1.0,1) \psline{-}(1.0,1)(1.5,1.5)
\psline{-}(1.5,1.5)(2.0,1) \psline{-}(2.0,1)(2.5,1.5)
\psline{-}(2.5,1.5)(3.0,1) \psline{-}(3.0,1)(3.5,1.5)

\psline{-}(6.5,1.5)(7.0,1) \psline{-}(7.0,1)(7.5,1.5)
\psline{-}(7.5,1.5)(8.0,2) \psline{-}(8.0,2)(8.5,1.5)
\psline{-}(8.5,1.5)(9.0,1) \psline{-}(9.0,1)(9.5,1.5)
}

{
\psset{linewidth=0.25pt,linestyle=dashed, dash=2.5pt 1.5pt,linecolor=black}
\psline{-}(7.5,1.5)(8,1) \psline{-}(8,1)(8.5,1.5)
}
{
\psset{dotsize=3.5pt}
\psdots(8.5,1.5)
\psset{fillcolor=white}
\psset{dotsize=4pt}\psset{dotstyle=o}
\psdots(8,2)
}
\end{pspicture} 

\begin{pspicture}(1,0)(9,2.5) 

{
\psset{linestyle=none}
\psset{fillstyle=solid}
\psset{fillcolor=lightgray}
\psframe(0.5,0.5)(3.5, 1)
\psframe(6.5,0.5)(9.5, 1)
\psframe(0.5,1.5)(3.5, 2)
\psframe(6.5,1.5)(9.5, 2)
}

{
\psset{linestyle=solid}
\psline{-}(0.5,0.5)(3.5,0.5)
\psline{-}(0.5,2)(3.5,2)
\psline{-}(0.5,0.5)(0.5,2)
\psline{-}(3.5,0.5)(3.5,2)

\psline{-}(6.5,0.5)(9.5,0.5)
\psline{-}(6.5,2)(9.5,2)
\psline{-}(6.5,0.5)(6.5,2)
\psline{-}(9.5,0.5)(9.5,2)

\psline{->}(4.5,1.25)(5.5,1.25) 
\rput(5.0,1.5){{\scriptsize $\cb _{x}$}}
}

{
\psset{linewidth=0.25pt,linestyle=dashed, dash=2.5pt 1.5pt,linecolor=gray}

\psline{-}(0.5,1.0)(3.5,1.0)  \psline{-}(6.5,1.0)(9.5,1.0)
\psline{-}(0.5,1.5)(3.5,1.5)  \psline{-}(6.5,1.5)(9.5,1.5)

\psline{-}(1.0,0.5)(1.0,2.0) \psline{-}(1.5,0.5)(1.5,2.0) 
\psline{-}(2.0,0.5)(2.0,2.0) \psline{-}(2.5,0.5)(2.5,2.0) 
\psline{-}(3.0,0.5)(3.0,2.0) \psline{-}(3.5,0.5)(3.5,2.0) 

\psline{-}(7.0,0.5)(7.0,2.0) \psline{-}(7.5,0.5)(7.5,2.0) 
\psline{-}(8.0,0.5)(8.0,2.0) \psline{-}(8.5,0.5)(8.5,2.0)
\psline{-}(9.0,0.5)(9.0,2.0) \psline{-}(9.5,0.5)(9.5,2.0)
}


\rput(0.25,1.0){{\scriptsize $l$}}
\rput(0.1,1.5){{\scriptsize $l+1$}}
\rput(-0.5,2){{\scriptsize ${\bf{c)}}$}}
\rput(9.75,1.0){{\scriptsize $l$}}
\rput(10,1.5){{\scriptsize $l+1$}}

\rput(2.0,0.25){{\scriptsize $x$}}
\rput(8.0,0.25){{\scriptsize $x$}}

{
\psset{linestyle=solid}

\psline{-}(0.5,1.5)(1.0,2) \psline{-}(1.0,2)(1.5,1.5)
\psline{-}(1.5,1.5)(2.0,2) \psline{-}(2.0,2)(2.5,1.5)
\psline{-}(2.5,1.5)(3.0,2) \psline{-}(3.0,2)(3.5,1.5)

\psline{-}(6.5,0.5)(7.0,1.0) \psline{-}(7.0,1.0)(7.5,0.5)
\psline{-}(7.5,0.5)(8.0,1.0) \psline{-}(8.0,1.0)(8.5,1.5)
\psline{-}(8.5,1.5)(9.0,2.0) \psline{-}(9.0,2.0)(9.5,1.5)
}

{
\psset{fillcolor=white}
\psset{dotsize=4pt}\psset{dotstyle=o}
\psdots(8.5,1.5)
}

\end{pspicture} 

\begin{pspicture}(1,0)(9,2.5) 

{
\psset{linestyle=none}
\psset{fillstyle=solid}
\psset{fillcolor=lightgray}
\psframe(0.5,0.5)(3.5, 1)
\psframe(6.5,0.5)(9.5, 1)
\psframe(0.5,1.5)(3.5, 2)
\psframe(6.5,1.5)(9.5, 2)
}

{
\psset{linestyle=solid}
\psline{-}(0.5,0.5)(3.5,0.5)
\psline{-}(0.5,2)(3.5,2)
\psline{-}(0.5,0.5)(0.5,2)
\psline{-}(3.5,0.5)(3.5,2)

\psline{-}(6.5,0.5)(9.5,0.5)
\psline{-}(6.5,2)(9.5,2)
\psline{-}(6.5,0.5)(6.5,2)
\psline{-}(9.5,0.5)(9.5,2)

\psline{->}(4.5,1.25)(5.5,1.25) 
\rput(5.0,1.5){{\scriptsize $\cbs _{x}$}}
}

{
\psset{linewidth=0.25pt,linestyle=dashed, dash=2.5pt 1.5pt,linecolor=gray}

\psline{-}(0.5,1.0)(3.5,1.0)  \psline{-}(6.5,1.0)(9.5,1.0)
\psline{-}(0.5,1.5)(3.5,1.5)  \psline{-}(6.5,1.5)(9.5,1.5)

\psline{-}(1.0,0.5)(1.0,2.0) \psline{-}(1.5,0.5)(1.5,2.0) 
\psline{-}(2.0,0.5)(2.0,2.0) \psline{-}(2.5,0.5)(2.5,2.0) 
\psline{-}(3.0,0.5)(3.0,2.0) \psline{-}(3.5,0.5)(3.5,2.0) 

\psline{-}(7.0,0.5)(7.0,2.0) \psline{-}(7.5,0.5)(7.5,2.0) 
\psline{-}(8.0,0.5)(8.0,2.0) \psline{-}(8.5,0.5)(8.5,2.0)
\psline{-}(9.0,0.5)(9.0,2.0) \psline{-}(9.5,0.5)(9.5,2.0)
}


\rput(0.25,1.0){{\scriptsize $l$}}
\rput(0.1,1.5){{\scriptsize $l+1$}}
\rput(-0.5,2){{\scriptsize ${\bf{d)}}$}}
\rput(9.75,1.0){{\scriptsize $l$}}
\rput(10,1.5){{\scriptsize $l+1$}}

\rput(2.0,0.25){{\scriptsize $x$}}
\rput(8.0,0.25){{\scriptsize $x$}}

{
\psset{linestyle=solid}

\psline{-}(0.5,1.0)(1.0,0.5) \psline{-}(1.0,0.5)(1.5,1.0)
\psline{-}(1.5,1.0)(2.0,0.5) \psline{-}(2.0,0.5)(2.5,1.0)
\psline{-}(2.5,1.0)(3.0,0.5) \psline{-}(3.0,0.5)(3.5,1.0)

\psline{-}(6.5,2)(7.0,1.5) \psline{-}(7.0,1.5)(7.5,2)
\psline{-}(7.5,2)(8.0,1.5) \psline{-}(8.0,1.5)(8.5,1.0)
\psline{-}(8.5,1.0)(9.0,0.5) \psline{-}(9.0,0.5)(9.5,1)
}

{
\psset{dotsize=3.5pt}
\psdots(8.5,1)
}

\end{pspicture} 

\end{center}
\end{figure}
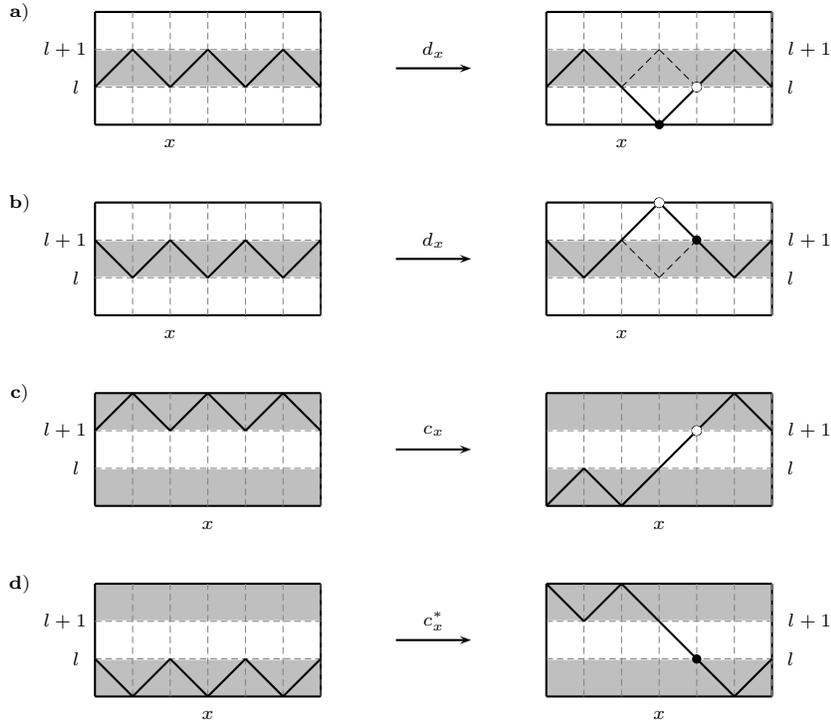

In the cases where $p'=2p+1$, and $a$ and $b$ are odd
integers with $1\le a<2p$ and $3\le b<2p$,
we encode $\vword{h}$ as follows.
After indexing the letters of the vertex word of $\vword{h}$ by
the $x$-coordinates of the corresponding vertices,
drop all letters $N$.
Then, reading from the left, replace each consecutive pair
$S_{x+1}S_{x+2}$ by $\db_x$.
Then replace each remaining $S_{y+1}$ by $\cb_y$ or $\cbs_y$ depending
on whether $y$ is even or odd.
This yields a finite sequence of \emph{symbols}
$\db_x$, $\cb_y$ and $\cbs_z$ which uniquely represents the
original path $h$.
We denote it $\oword{h}$ and refer to it as the
\emph{operator word} of $h$.
In the case of the path $h$ of Fig.~\ref{TypicalScoringFig},
from \eqref{Eq:TypicalVword}, we obtain
\begin{equation}\label{Eq:TypicalOword}
\oword{h}=\cb_0\,\db_2\,\cb_4\,\cb_6\,\db_{10}\,\cbs_{13}
          \,\db_{16}\,\db_{19}\,\cbs_{21}\,.
\end{equation}
Each symbol $\db_x$, $\cb_y$ or $\cbs_z$ may be viewed as
an operator which changes the
nature of three or two vertices of an oscillating
portion of a path near position $x$, $y$ or $z$ respectively.
These actions are illustrated in Fig.~\ref{Fig:CCDaction}.
The path $h$ then results from the action of the operator word
$\oword{h}$ on the purely oscillating
\emph{vacuum path} $h^{\text{vac$(b)$}}\in\PP^p_{b,b}$,
defined by $h^{\text{vac$(b)$}}_x=b$ for $x$ even
and $h^{\text{vac$(b)$}}_x=b-1$ for $x$ odd.

Note that the operator $\db_x$ acts locally to generate either a
peak or a valley in a light band, changing neither the path's
startpoint nor tail.
In contrast, the actions of the operators $\cb_x$ and $\cbs_x$ are
non-local
in the sense that they change the
portion of the path lying between 0 and $x+1$.
In particular, they change the startpoint of the path,
decreasing and increasing it by 2 respectively.%
\footnote{
  The operators $\cb_x$ and $\cbs_x$ map between
  paths which label states of different modules.
  Their actions could be defined
  either to maintain the tail and change the initial point
  ($\PP^p_{a,b}\rw \PP^p_{a\mp2,b}$), as described in the main
  text, or, alternatively, to change the tail and maintain the initial
  point ($\PP^p_{a,b}\rw \PP^p_{a,b\pm2}$), as in \cite{PMnlob}
  for unitary RSOS paths.
  Under the action of $\cb^{(*)}_x$, the paths are then modified
  either in the interval $[0,x+1]$,
  as in the main text,
  or, alternatively, in the interval $[x,\y]$.
  Either choice corresponds to a genuine non-local action.
  However, the first one is somewhat more localized.
  This is one motivation for the choice made here.
  But the ultimate reason lies in the expected greater simplicity
  of this choice in the treatment of other (than the $p'=2p+1$)
  classes of models.
}

Therefore, because the operator word $\oword{h}$ maps between elements
of $\PP^p_{b,b}$ and $\PP^p_{a,b}$,
it follows that the difference between the number of operators
$\cb_y$ and $\cbs_z$ in the operator word $\oword{h}$ is $(b-a)/2$.

Obtained as described above, the subscripts of neighbouring pairs
of operators in $\oword{h}$ naturally satisfy certain constraints.
They are:
\begin{equation}\label{Eq:RSOSconstraints}
\begin{split}
\db_{x}\:\cb^{(*)}_{x'} ,\,
\cb^{(*)}_{x}\:\db_{x'},\,
\db_x \:\db_{x'},\,
\cb_{x}\:\cb_{x'},\,
\cb^{*}_{x}\:\cb^{*}_{x'}
&\quad\implies\quad x'\ge x+2,\\
\cb_{x} \: \cbs_{x'},\,
\cbs_{x}\:\cb_{x'}
&\quad\implies\quad x'\ge x+3,
\end{split}
\end{equation}
where $\cb^{(*)}$ denotes either $\cb$ or $\cb^*$.
%
%
A word $\osymbol$ in the symbols $\db_x$, $\cb_y$, $\cbs_z$ is
called standard if every neighbouring pair in $\osymbol$ respects
the constraints \eqref{Eq:RSOSconstraints}.
 
In the operator word $\oword{h}$, the particles described in
Section \ref{Sec:RemFB} correspond to the symbols $d_x$.
The excitation of each such particle is given by the
number of non-scoring vertices in $h$ to its left.
Therefore, for the $i$th operator $\db_x$ in $\oword{h}$,
counted from the right, this excitation is given by
\begin{equation}\label{Eq:RSOSexcite}
\lambda_i=
x-2\#\{{\db}\cdots\db_x\}-\#\{{\cb^{(*)}}\cdots\db_x\},
\end{equation}
where $\#\{A\cdots B\}$ denotes the number of pairs $A$ and $B$
of operators in $\oword{h}$ with $A$ to the left of $B$.
For instance, consider the excitation $\lambda_1$ of the particle
corresponding to $d_{19}$ in the word \eqref{Eq:TypicalOword}:
there are three pairs of the first type:
$(d_2,d_{19}),\, (d_{10},d_{19})$, and $(d_{16},d_{19})$,
and four pairs of the second type:
$(c_0,d_{19}),\, (c_{4 },d_{19}),\, (c_6,d_{19})$, and
$(c^*_{13},d_{19})$.
Thus, in this case, the excitation is $\lambda_1=19-6-4=9$.

For an arbitrary word $\osymbol$ in the symbols
$\db_x$, $\cb_y$, $\cbs_z$, we use \eqref{Eq:RSOSexcite} to define
$\lambda_i$ for the $i$th operator $\db_x$,
counted from the right.
Then define
the vector $\lambda(\osymbol)=(\lambda_1,\lambda_2,\cdots,\lambda_n)$,
where $n$ is the number of operators $\db_x$ in $\osymbol$.
If $\lambda_1\ge\lambda_2\ge\cdots\ge\lambda_n\ge0$
(so that $\lambda(\osymbol)$ is a partition),
then we say that $\osymbol$ is \emph{physical}.
Otherwise, we say that $\osymbol$ is \emph{unphysical}.
The constraints \eqref{Eq:RSOSconstraints}
guarantee that every standard word $\oword{h}$ is physical.

We now consider the effect on $\oword{h}$ of changing the
excitations of the particles in $h$.
First consider incrementing or decrementing the excitation of the
particle corresponding to an operator $\db_x$.
This is possible if and only if we obtain a physical word $\osymbol'$
on replacing the operator $\db_x$ in $\oword{h}$ by $\db_{x+1}$
or $\db_{x-1}$ respectively
(otherwise, the particle is `blocked':
two particles cannot occupy the same position).
Even if physical, the word $\osymbol'$ might not be standard.
However, if standard, the resulting path 
is readily obtained from $\osymbol'$ via the
actions given in Fig.~\ref{Fig:CCDaction}.
If it is not standard, then necessarily we would have made
one of the local changes:
\begin{subequations}\label{Eq:ToViolate}
\begin{align}
\label{Eq:ToViolate1}
\db_x\, \cb^{(*)}_{x+2}\, &\to\, \db_{x+1}\, \cb^{(*)}_{x+2},\\
\label{Eq:ToViolate2}
\cb^{(*)}_{x-2}\, \db_x\, &\to\,  \cb^{(*)}_{x-2}\, \db_{x-1},
\end{align}
\end{subequations}
depending on whether we incremented or decremented the excitation.
These violations of \eqref{Eq:RSOSconstraints}
may be seen to arise because, within an odd length
sequence of scoring vertices, there is
ambiguity over which pairs correspond to the particles.
For example, the sequence $NSSSN$ might be interpreted with
the first two $S$s being the particle, thereby yielding the operators
$\db_1\,\cb_3$, or the latter two $S$s being the particle,
thereby yielding the operators $\cb_1\,\db_2$.
Thus, we should impose the equivalence
\begin{equation}\label{Eq:SimpleEquiv}
\db_x\, \cb^{(*)}_{x+2}\, \equiv\, \cb^{(*)}_x\, \db_{x+1}.
\end{equation}
In the case of the moves \eqref{Eq:ToViolate}, this equivalence
should be applied before increasing the excitation in the
case \eqref{Eq:ToViolate1}, and after decreasing the excitation
in the case \eqref{Eq:ToViolate2}.  

To excite the particle corresponding to a particular $\db_x$ by
more than 1, we can proceed as above, one step at a time.
However, we easily see that the process can be streamlined
by first adding the required excitement to the subscript of $d_x$
(this excitation being possible if and only if the resulting
word is physical),
and then for each pair $\db_y\,{\cb^{(*)}_{y'}}$ with $y'<y+2$,
imposing the equivalence
\begin{equation}\label{Eq:RSOSEquiv}
\db_y\, \cb^{(*)}_{y'}\, \equiv\, \cb^{(*)}_{y'-2}\, \db_{y+1}.
\end{equation}
For example, consider the operator word
\begin{equation}\label{Eq:B2word}
\db_0\,\db_2\,\db_4\,\db_6\,
\cb_{8}\,\cb_{10}\,\cb_{12}\,\cbs_{17}\,\cbs_{21}.
\end{equation}
Exciting the rightmost 
particle (represented by $\db_6$) by 9 yields first
\begin{equation}
\db_0\,\db_2\,\db_4\,\db_{15}\,
\cb_{8}\,\cb_{10}\,\cb_{12}\,\cbs_{17}\,\cbs_{21}.
\end{equation}
Repeatedly applying \eqref{Eq:RSOSEquiv} yields the following
sequence of words:
\begin{equation}
\begin{split}
\db_0\,\db_2\,\db_4\,(\db_{15}\,
\cb_{8})\,\cb_{10}\,\cb_{12}\,\cbs_{17}\,\cbs_{21}
&\quad\to\quad
\db_0\,\db_2\,\db_4\,\cb_{6}\,
(\db_{16}\,\cb_{10})\,\cb_{12}\,\cbs_{17}\,\cbs_{21}\\
&\quad\to\quad
\db_0\,\db_2\,\db_4\,\cb_{6}\,
\cb_{8}\,(\db_{17}\,\cb_{12})\,\cbs_{17}\,\cbs_{21}\\
&\quad\to\quad
\db_0\,\db_2\,\db_4\,\cb_{6}\,
\cb_{8}\,\cb_{10}\,(\db_{18}\,\cbs_{17})\,\cbs_{21}\\
&\quad\to\quad
\db_0\,\db_2\,\db_4\,\cb_{6}\,
\cb_{8}\,\cb_{10}\,\cbs_{15}\,\db_{19}\,\cbs_{21},
\end{split}
\end{equation}
where, in each line, we have used parentheses to indicate
the pair of operators affected.
Conversely, we can reduce the excitation of a $d_x$ by decreasing
its subscript by the required amount
(again, this is possible if and only if the resulting
word is physical),
and then repeatedly using
\eqref{Eq:RSOSEquiv} in the cases $y'\ge y+2$
to reexpress the right side as the left side.
In either case, once the standard operator word has been obtained,
the corresponding RSOS path can be readily constructed using
Fig.~\ref{Fig:CCDaction}.

The construction may be applied to excite a number of
particles simultaneously.
For example, applying the excitations 1, 5, 8 and 9 to the particles
of the operator word \eqref{Eq:B2word} yields the non-standard
word
\begin{equation}
\db_{1}\,\db_{7}\,\db_{12}\,\db_{15}\,
\cb_{8}\,\cb_{10}\,\cb_{12}\,\cbs_{17}\,\cbs_{21}.
\end{equation}
It may be checked that repeated use of \eqref{Eq:RSOSEquiv} then
yields the standard word given in \eqref{Eq:TypicalOword},
which corresponds to the path of Fig.~\ref{TypicalScoringFig}.%
\footnote{Operator words of the form \eqref{Eq:B2word} which
  begin with the $n$ operators $\db_0\,\db_2\,\db_4\,\cdots\,\db_{2n-2}$
  are, in the language of \cite{FLPW}, the result of a
  $\BB_2(n)$-transform on a particle-deficient path.
  The process of exciting these particles
  corresponds to the $\BB_3(\lambda)$-transform, 
  with the excitation of the $i$th particle,
  {counted from the right,} 
  specified by the part
  $\lambda_{i}$ of the partition $\lambda$.}

Finally, for this section, we note that the excitation of a
particle corresponding to an operator $\db_x$
in an operator word $\osymbol$ is given by
\eqref{Eq:RSOSexcite}, even when the word is non-standard.
This is so because the value of the expression
\eqref{Eq:RSOSexcite} is unchanged for operator words obtained
from one another through the equivalence \eqref{Eq:RSOSEquiv}.

\subsection{$\HH^p$ paths as sequences of operators}
\label{Sec:HALFop}

Here, we provide an encoding of the half-lattice paths of
$\HH^p_{\hat a,\hat b}$ that is analogous to that of the
previous section.

For $\hat a,\hat b\in\ZZ$, let $\hh\in\HH^p_{\hat a,\hat b}$.
Because $\hh$ has no peak at a non-integer height, it follows
that if there is a straight-up vertex at position $x\in\ZZ$,
there must also be one at position $x+1/2$,
and if there is a straight-down vertex at position $x\in\ZZph$,
there must either be a straight-down vertex at position $x+1/2$
or a straight-up vertex at position $x+1$.
Similarly, if there is a straight-down vertex at position $x\in\ZZ$,
there must also be one at position $x-1/2$,
and if there is a straight-up vertex at position $x\in\ZZph$,
there must either be a straight-up vertex at position $x-1/2$
or a straight-down vertex at position $x-1$.
Consequently, working from the left of $\hh$, we may consecutively pair
the straight vertices such that each pair occurs at neighbouring
positions $x$ and $x+1/2$, or next neighbouring positions
$x$ and $x+1$
(through the convention stated in Section \ref{Sec:Halfpath},
the vertex at $(0,\hat a)$ is deemed straight if and only if
the first segment of $\hh$ is in the NE direction).
We encode each of those pairs that occur at positions $x$ and $x+1/2$
using $\ch_x$ or $\chs_x$ depending on whether $x$ is
integer or non-integer respectively.
In the other case, where the pair occurs at positions $x$ and $x+1$,
we encode the pair using $\dh_x$.
In this latter case, note that $x\in\ZZph$ and that there is
a valley at the intermediate position $x+1/2$.
%
Let $\owordh{\hh}$ denote the word in the symbols
$\dh_x$, $\ch_y$, $\chs_z$ obtained in this way,
ordered with increasing subscripts.
We refer to it as the \emph{operator word} of $\hh$.
Of course, the half-lattice path $\hh$ can be immediately
recovered from $\owordh{\hh}$.%
\footnote{
  As mentioned in the introduction of the section,
  this operator construction of the $\HH^p$ paths is a modified
  version of the one already presented in \cite{PMnlob}.
  To substantiate this statement, let
  us recall briefly the operator construction of $\HH^p$
  paths in \cite{PMnlob}.
  The paths are constructed from the appropriate ground state
  by the action of a sequence of non-local operators $b_x,\, b_x^* $:
  $b_x$ transforms a peak at $x$ into a straight-up segment
  and $b_x^*$ transforms a valley at $x$ into a straight-down segment
  (both segments linking the points $x$ and $x+1/2$).  
  These actions modify the path for all $x'\geq x+1/2$
  (in contradistinction with the actions of the operators
  defined in the main text, which are limited to the initial portion
  of the path, namely $x'\leq x$).
  The constraint on the integrality of the peak positions shows
  that the operators $b$ and $b^*$ always occur in successive pairs
  that are either of the types $(b \, b)$, $(b^* \, b^*)$ or $(b^* \, b)$,
  with subindices differing by a half-integer in the first two cases and
  by an integer in the third one.
  In terms of these operators,
  the operators $ \tilde c, \,\tilde c^*$ and $\dh$
  are expressed
  $$
  \tilde c_x = b_x \: b_{x+1/2},\quad
  \tilde c^*_{x'} = b^{*}_{x'} \: b^{*}_{x'+1/2},\quad
  \dh_{x'} = b^{*}_{x'} \: b_{x'+1}.
  $$
  $\dh $ is the same as the operator defined in the main text,
  while $\tilde c$ and $\tilde c^*$ are roughly the operators
  $\ch $ and $\chs$ but acting
  instead to maintain the path's initial portion, while changing
  its tail. 
  Summing up, our present operator construction is different from
  that introduced in \cite{PMnlob} in two ways:
  it uses a reduced number of operators and the operator action
  modifies the initial rather than the final portion of the path.
  Let us now turn to the
  interpretation where particles are viewed as the path's basic
  constituents \cite{OleJS,PMnlob,PMprsos}.
  Given that every path is described by a sequence composed of
  an equal number of operators $b$ and $b^*$, it can be seen, from the
  decomposition of a path into charged peaks \cite{PMnpd}, that there
  are actually $2 p-3$ combinations of the $b$ and $b^*$ that are allowed.
  These are the $l$-blocks --- the particles whose numbers are the
  summation variables in the fermionic character --- defined by \cite{PMnlob}:
  $$
  b^{l-1} \: b^{*l} \: b \quad  \text{for $l$ odd} \quad \text{and}\quad
  b^l \:  b^{*l}  \quad \text{for $l$ even},\quad \text{
  where}\quad 1\leq l \leq 2 {{ p}}  - 3.$$
  In terms of the new operators, the path building blocks are seen
  to be simply the $p-1$ combinations:
  $\dh $ and $ {\tilde c}^{\ell} \: {\tilde c}^{*\ell}$, or equivalently
  $\ch^\ell\ch^{*\ell}$, for $1\leq \ell\leq p-2$.
  (In the terminology of \cite{PMprsos}, these particles are interpreted
   as a breather and pairs of kinks-antikinks of topological
   charge $\ell$ respectively.)
  This particle interpretation is the starting point for a
  direct derivation of the characters
  (in the line of \cite{OleJS}) that matches the usual expressions
  given in \cite{Melzfermionic,BMlmp,FQ,FLPW,TWfe,PMprsos}.
  }


To illustrate this construction,
consider the path $\hh$ of Fig.~\ref{che_h}.
Here, the operator word $\owordh{\hh}$ is found to be
\begin{equation} \label{Eq:seq_cced}
\hat{\pi}(h) \, = \, \ch_{0} \: \ch_{1} \: \ch_{2} \: \chs_{\frac{9}{2}} \:
\dh_{\frac{11}{2}} \: \chs_{\frac{15}{2}} \: \dh_{\frac{25}{2}} \:
\dh_{\frac{35}{2}} \: \dh_{\frac{41}{2}}. 
\end{equation}

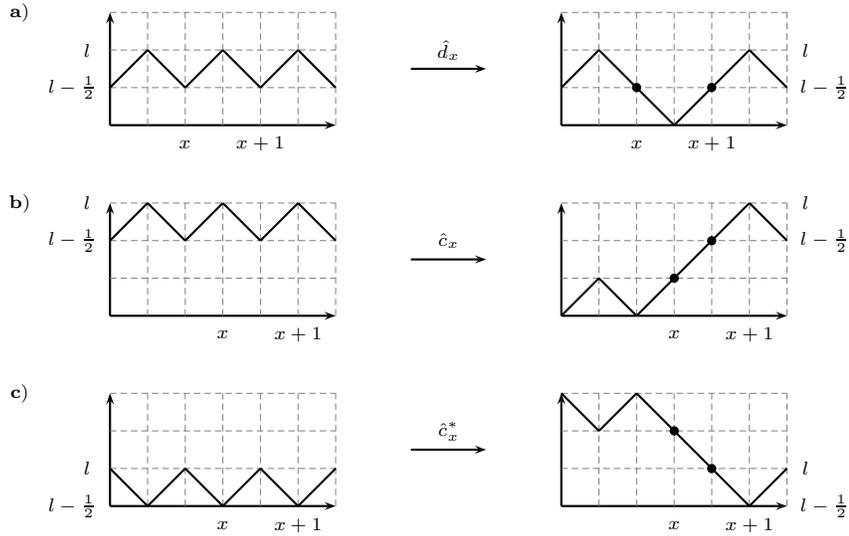
\begin{figure}[ht]
\caption{{\footnotesize The action of the operators $\dh_x$, $\ch_x$
  and $\chs_x$ on a portion of a path that oscillates between
  heights $l$ and $l-1/2$, where $l\in\ZZ$.
  The black dots are the vertices that contribute to the weight.}}
\vskip-1.4cm
\label{Fig:CCDhaction}
\begin{center}
\begin{pspicture}(1,0)(9,3.5)  

{
\psset{linestyle=solid}
\psline{->}(0.5,0.5)(3.5,0.5)
\psline{->}(0.5,0.5)(0.5,2)

\psline{->}(6.5,0.5)(9.5,0.5)
\psline{->}(6.5,0.5)(6.5,2)

\psline{->}(4.5,1.25)(5.5,1.25) 
\rput(5.0,1.5){{\scriptsize $\dh _{x}$}}
}

{
\psset{linewidth=0.25pt,linestyle=dashed, dash=2.5pt 1.5pt,linecolor=gray}

\psline{-}(0.5,1.0)(3.5,1.0)  \psline{-}(6.5,1.0)(9.5,1.0)
\psline{-}(0.5,1.5)(3.5,1.5)  \psline{-}(6.5,1.5)(9.5,1.5)
\psline{-}(0.5,2)(3.5,2)      \psline{-}(6.5,2)(9.5,2)

\psline{-}(1.0,0.5)(1.0,2.0) \psline{-}(1.5,0.5)(1.5,2.0) 
\psline{-}(2.0,0.5)(2.0,2.0) \psline{-}(2.5,0.5)(2.5,2.0) 
\psline{-}(3.0,0.5)(3.0,2.0) \psline{-}(3.5,0.5)(3.5,2.0) 

\psline{-}(7.0,0.5)(7.0,2.0) \psline{-}(7.5,0.5)(7.5,2.0) 
\psline{-}(8.0,0.5)(8.0,2.0) \psline{-}(8.5,0.5)(8.5,2.0)
\psline{-}(9.0,0.5)(9.0,2.0) \psline{-}(9.5,0.5)(9.5,2.0)
}


\rput(-0.7,2){{\scriptsize ${\bf{a)}}$}}
\rput(0.2,1.5){{\scriptsize $l$}}
\rput(0.0,1.0){{\scriptsize $l-\frac{1}{2}$}}
\rput(9.75,1.5){{\scriptsize $l$}}
\rput(10,1.0){{\scriptsize $l-\frac{1}{2}$}}

\rput(1.5,0.25){{\scriptsize $x$}} \rput(2.5,0.25){{\scriptsize $x+1$}}
\rput(7.5,0.25){{\scriptsize $x$}} \rput(8.5,0.25){{\scriptsize $x+1$}}

{
\psset{linestyle=solid}

\psline{-}(.5,1)(1.0,1.5) \psline{-}(1.0,1.5)(1.5,1)
\psline{-}(1.5,1)(2.0,1.5) \psline{-}(2.0,1.5)(2.5,1)
\psline{-}(2.5,1)(3.0,1.5) \psline{-}(3.0,1.5)(3.5,1)

\psline{-}(6.5,1)(7.0,1.5) \psline{-}(7.0,1.5)(7.5,1.0)
\psline{-}(7.5,1.0)(8.0,0.5) \psline{-}(8.0,0.5)(8.5,1)
\psline{-}(8.5,1)(9.0,1.5) \psline{-}(9.0,1.5)(9.5,1)
}

{
\psset{dotsize=3.5pt}
\psdots(7.5,1)(8.5,1)
}

\end{pspicture}   

\begin{pspicture}(1,0)(9,2.5)  

{
\psset{linestyle=solid}
\psline{->}(0.5,0.5)(3.5,0.5)
\psline{->}(0.5,0.5)(0.5,2)

\psline{->}(6.5,0.5)(9.5,0.5)
\psline{->}(6.5,0.5)(6.5,2)

\psline{->}(4.5,1.25)(5.5,1.25) 
\rput(5.0,1.5){{\scriptsize $\ch _{x}$}}
}

{
\psset{linewidth=0.25pt,linestyle=dashed, dash=2.5pt 1.5pt,linecolor=gray}

\psline{-}(0.5,1.0)(3.5,1.0)  \psline{-}(6.5,1.0)(9.5,1.0)
\psline{-}(0.5,1.5)(3.5,1.5)  \psline{-}(6.5,1.5)(9.5,1.5)
\psline{-}(0.5,2)(3.5,2)      \psline{-}(6.5,2)(9.5,2)
 
\psline{-}(1.0,0.5)(1.0,2.0) \psline{-}(1.5,0.5)(1.5,2.0) 
\psline{-}(2.0,0.5)(2.0,2.0) \psline{-}(2.5,0.5)(2.5,2.0) 
\psline{-}(3.0,0.5)(3.0,2.0) \psline{-}(3.5,0.5)(3.5,2.0) 

\psline{-}(7.0,0.5)(7.0,2.0) \psline{-}(7.5,0.5)(7.5,2.0) 
\psline{-}(8.0,0.5)(8.0,2.0) \psline{-}(8.5,0.5)(8.5,2.0)
\psline{-}(9.0,0.5)(9.0,2.0) \psline{-}(9.5,0.5)(9.5,2.0)
}


\rput(-0.7,2){{\scriptsize ${\bf{b)}}$}}
\rput(0.2,2.0){{\scriptsize $l$}}
\rput(0.0,1.5){{\scriptsize $l-\frac{1}{2}$}}
\rput(9.75,2.0){{\scriptsize $l$}}
\rput(10,1.5){{\scriptsize $l-\frac{1}{2}$}}

\rput(2.0,0.25){{\scriptsize $x$}} \rput(3.0,0.25){{\scriptsize $x+1$}}
\rput(8.0,0.25){{\scriptsize $x$}} \rput(9.0,0.25){{\scriptsize $x+1$}}

{
\psset{linestyle=solid}

\psline{-}(0.5,1.5)(1.0,2.0) \psline{-}(1.0,2.0)(1.5,1.5)
\psline{-}(1.5,1.5)(2.0,2.0) \psline{-}(2.0,2.0)(2.5,1.5)
\psline{-}(2.5,1.5)(3.0,2.0) \psline{-}(3.0,2.0)(3.5,1.5)

\psline{-}(6.5,0.5)(7.0,1.0) \psline{-}(7.0,1.0)(7.5,0.5)
\psline{-}(7.5,0.5)(8.0,1.0) \psline{-}(8.0,1.0)(8.5,1.5)
\psline{-}(8.5,1.5)(9.0,2.0) \psline{-}(9.0,2.0)(9.5,1.5)
}

{
\psset{dotsize=3.5pt}
\psdots(8,1)(8.5,1.5)
}

\end{pspicture}   

\begin{pspicture}(1,0)(9,2.5)  

{
\psset{linestyle=solid}
\psline{->}(0.5,0.5)(3.5,0.5)
\psline{->}(0.5,0.5)(0.5,2)

\psline{->}(6.5,0.5)(9.5,0.5)
\psline{->}(6.5,0.5)(6.5,2)

\psline{->}(4.5,1.25)(5.5,1.25) 
\rput(5.0,1.5){{\scriptsize $\chs _{x}$}}
}

{
\psset{linewidth=0.25pt,linestyle=dashed, dash=2.5pt 1.5pt,linecolor=gray}

\psline{-}(0.5,1.0)(3.5,1.0)  \psline{-}(6.5,1.0)(9.5,1.0)
\psline{-}(0.5,1.5)(3.5,1.5)  \psline{-}(6.5,1.5)(9.5,1.5)
\psline{-}(0.5,2)(3.5,2)      \psline{-}(6.5,2)(9.5,2)

\psline{-}(1.0,0.5)(1.0,2.0) \psline{-}(1.5,0.5)(1.5,2.0) 
\psline{-}(2.0,0.5)(2.0,2.0) \psline{-}(2.5,0.5)(2.5,2.0) 
\psline{-}(3.0,0.5)(3.0,2.0) \psline{-}(3.5,0.5)(3.5,2.0) 

\psline{-}(7.0,0.5)(7.0,2.0) \psline{-}(7.5,0.5)(7.5,2.0) 
\psline{-}(8.0,0.5)(8.0,2.0) \psline{-}(8.5,0.5)(8.5,2.0)
\psline{-}(9.0,0.5)(9.0,2.0) \psline{-}(9.5,0.5)(9.5,2.0)
}


\rput(-0.7,2){{\scriptsize ${\bf{c)}}$}}
\rput(0.2,1.0){{\scriptsize $l$}}
\rput(0.0,0.5){{\scriptsize $l-\frac{1}{2}$}}
\rput(9.75,1.0){{\scriptsize $l$}}
\rput(10,0.5){{\scriptsize $l-\frac{1}{2}$}}

\rput(2.0,0.25){{\scriptsize $x$}} \rput(3.0,0.25){{\scriptsize $x+1$}}
\rput(8.0,0.25){{\scriptsize $x$}} \rput(9.0,0.25){{\scriptsize $x+1$}}

{
\psset{linestyle=solid}

\psline{-}(0.5,1)(1.0,0.5) \psline{-}(1.0,0.5)(1.5,1)
\psline{-}(1.5,1)(2.0,0.5) \psline{-}(2.0,0.5)(2.5,1)
\psline{-}(2.5,1)(3.0,0.5) \psline{-}(3.0,0.5)(3.5,1)

\psline{-}(6.5,2)(7.0,1.5) \psline{-}(7.0,1.5)(7.5,2)
\psline{-}(7.5,2)(8.0,1.5) \psline{-}(8.0,1.5)(8.5,1.0)
\psline{-}(8.5,1.0)(9.0,0.5) \psline{-}(9.0,0.5)(9.5,1)
}
{
\psset{dotsize=3.5pt}
\psdots(8,1.5)(8.5,1)
}

\end{pspicture} 

\end{center}
\end{figure}

Each symbol $\dh_x$, $\ch_y$ or $\chs_z$ may be viewed as an operator
which changes the nature of two vertices of an oscillating
portion of a path near position $x$, $y$ or $z$ respectively.
These actions are illustrated in Fig.~\ref{Fig:CCDhaction}.
The path $\hh$ then results from the action of the operator word
$\owordh{\hh}$ on
the purely oscillating \emph{vacuum path}
$\hh^{\text{vac$(\hat b)$}}\in\HH^p_{\hat b,\hat b}$, defined by
$\hh^{\text{vac$(\hat b)$}}_x=\hat b$
and $\hh^{\text{vac$(\hat b)$}}_{x+1/2}=\hat b-1/2$ for
$x\in\ZZ_{\ge0}$.


Note that the actions of the operators $\ch_y$ and $\chs_z$ are
non-local 
in the sense that they change the starting height of the path,
decreasing and increasing it by 1 respectively.
In contrast, $\dh_x$ does not affect the path's starting point,
acting locally to change an integer peak into an integer valley.  

The subscripts of the symbols in the word $\owordh{\hh}$ naturally
satisfy certain constraints. They are:
%
%
%
\begin{equation}\label{Eq:HALFconstraints}
\begin{split}
\ch_{x}\:\ch_{x'},\,
\ch^{*}_{x}\:\ch^{*}_{x'},\,
\ch^{*}_{x}\:\db_{x'}
&\quad\implies\quad x'\ge x+1,\\
\ch^{*}_{x}\:\ch_{x'},\,
\ch_{x}\:\ch^{*}_{x'},\,
\dh_{x}\:\ch_{x'},\,
\ch_{x}\:\dh_{x'}
&\quad\implies\quad x'\ge x+\frac32,\\
\dh_{x}\:\dh_{x'}\,,\,
\dh_{x}\:\chs_{x'}
&\quad\implies\quad x'\ge x+2.
\end{split}
\end{equation}
Later, we consider arbitrary words in the operators
$\dh_x$, $\ch_y$ and $\chs_z$, with each $y\in\ZZ$ and each $x,z\in\ZZph$.
Such a word $\osymbolh$ is called \emph{standard}
if every neighbouring pair in $\osymbolh$ respects the
constraints \eqref{Eq:HALFconstraints}.


Each operator $\dh_x$ in $\owordh{\hh}$ corresponds to an integer
valley of $\hh$ at position $x+1/2\in\ZZ$.
We define the excitation of this integer valley to be
the number of non-integer valleys to its left.
To express this excitation in terms of the
operator word $\owordh{\hh}$,
%
%
let $\#\{A\cdots B\}$ denote the number of pairs $A$ and $B$
of operators in $\owordh{\hh}$ with $A$ to the left of $B$.
The number of straight vertices strictly to the left of position
$x+1$ is then $2\#\{\dh\cdots\dh_x\}+2\#\{\ch^{(*)}\cdots\dh_x\}+1$,  
where $\ch^{(*)}$ denotes either $\ch$ or $\ch^*$
(the $+1$ accounting for the straight-down vertex of $\dh_x$ at $x$).
Thus, the number of valleys and peaks strictly to the left of
$x+1$ is $2x+1-2\#\{\dh\cdots\dh_x\}-2\#\{\ch^{(*)}\cdots\dh_x\}$.
Exactly half of these are valleys.
Of those, $\#\{\dh\cdots\dh_x\}+1$ are integer valleys.
Therefore, the excitation of the integer valley corresponding to the
$i$th operator $\dh_x$ in $\owordh{\hh}$,
counted from the right, is given by
\begin{equation}\label{Eq:HALFexcite}
\hat\lambda_i={}
x-\frac12-2\#\{\dh\cdots\dh_x\}-\#\{\ch^{(*)}\cdots\dh_x\}.
\end{equation}

Proceeding as in Section \ref{Sec:RSOSop},
for an arbitrary word $\osymbolh$ in the symbols
$\dh_x$, $\ch_y$, $\chs_z$, we use \eqref{Eq:HALFexcite} to define
$\hat\lambda_i$ for the $i$th operator $\dh_x$,
counted from the right.
Then define
the vector $\hat\lambda(\osymbolh)
                  =(\hat\lambda_1,\hat\lambda_2,\cdots,\hat\lambda_n)$,
where $n$ is the number of operators $\dh_x$ in $\osymbolh$.
If $\hat\lambda_1\ge\hat\lambda_2\ge\cdots\ge\hat\lambda_n\ge0$
then we say that $\osymbolh$ is \emph{physical},
and \emph{unphysical} otherwise.
Again, the constraints \eqref{Eq:HALFconstraints}
ensure that every standard word $\owordh{\hh}$ is physical.


We now consider the effect on $\owordh{\hh}$ of changing the
excitations of the particles in $\hh$,
mirroring 
the analysis of the previous section with some adjustments.
First consider incrementing the excitation
of the integer valley corresponding to an operator $\dh_x$.
This is possible if and only if we obtain a physical word $\osymbolh'$
on replacing the operator $\dh_x$ in $\owordh{\hh}$ by $\dh_{x+1}$
(otherwise, the integer valley is `blocked').
Here again, even if physical, the word $\osymbolh'$ 
might not be standard.

The effect on $\hh$ of incrementing the excitation is to make shallow
the valley of $\hh$ at position $x+1/2$, and to deepen the subsequent
valley.
Let $\hh'$ be the resulting path.
We claim that if $\osymbolh'$ 
is standard then $\owordh{\hh'}=\osymbolh'$.
This is so because, if $\osymbolh'$ is standard then, in $\owordh{\hh}$,
the operator $\dh_x$ is immediately followed by either an operator
$\ch^{(*)}_{x'}$ or an operator $\dh_{x'}$ (or nothing) with
$x'>x+2$.
Thus, the subsequent valley to be deepened is at position $x+2$.
We then see that $\owordh{\hh'}=\osymbolh'$, as claimed.

On the other hand, if $x'\le x+2$,
the word $\osymbolh'$ is non-standard.
In this case, we ascertain the position of the subsequent valley
in $\hh$ from the sequence of operators that follow
$\dh_x$ in $\owordh{\hh}$.
In this word, $\dh_x$ and subsequent operators
necessarily form the standard subword
\begin{equation}\label{Eq:HalfSeq1}
\dh_x\, \ch_{x+3/2}\, \ch_{x+5/2} \cdots \ch_{x+t+1/2}\,
\chs_{x+t+2}\, \chs_{x+t+3} \cdots \chs_{x+t+t^*+1},
\end{equation}
where $t\ge0$, $t^*\ge0$, with at least one of these non-zero,
and any subsequent operators have subscripts at least $x+t+t^*+5/2$
(there cannot be a subsequent operator $\dh_{x+t+t^*+2}$ because,
as is readily checked using \eqref{Eq:HALFexcite},
with such an operator, the word {$\osymbolh'$} 
would be unphysical).
The subsequent valley is then at position $x+t+t'+2$.
Therefore, the word $\owordh{\hh'}$ is obtained from
$\owordh{\hh}$ by replacing the subword \eqref{Eq:HalfSeq1} by
\begin{equation}\label{Eq:HalfSeq3}
\ch_{x+1/2}\, \ch_{x+3/2} \cdots \ch_{x+t-1/2}\,
\chs_{x+t+1}\, \chs_{x+t+2} \cdots \chs_{x+t+t^*}\, \dh_{x+t+t^*+1}. 
\end{equation}

Alternatively, we may proceed in a similar way to that in
Section \ref{Sec:RSOSop}, using non-standard words.
Then, to effect the excitement, first increment the subscript
of $\dh_x$ in $\owordh{\hh}$, regardless of subsequent operators.
The excitement is possible if the resulting word $\osymbolh'$
is physical.
Then, if the word is standard, it is $\owordh{\hh'}$.
Otherwise, a subword of the form
\eqref{Eq:HalfSeq1} must have been present in $\owordh{\hh}$.
We then proceed by imposing the equalities
\begin{subequations}\label{Eq:HalfEquiv0}
\begin{align}
\label{Eq:HalfEquiv1}
\dh_y\, \ch_{y+1/2}\, &\equiv\, \ch_{y-1/2}\, \dh_{y+1},\\
\label{Eq:HalfEquiv2}
{\dh_y}\, \ch^{*}_{y+1}\, &\equiv\, \ch^{*}_y\, \dh_{y+1},
\end{align}
\end{subequations}
until a standard word results.
This standard word is $\owordh{\hh'}$ because,
after replacing $\dh_x$ by $\dh_{x+1}$ in \eqref{Eq:HalfSeq1},
this procedure produces \eqref{Eq:HalfSeq3}.

As in Section \ref{Sec:RSOSop}, we can streamline the process of
exciting a particular $\dh_x$, by adding the required excitation
to the subscript
(this excitation being possible if and only if the resulting
word is physical)
and then repeatedly imposing
\begin{equation}
\label{Eq:HalfEquiv}
\dh_y\, \ch^{(*)}_{y'}\, \equiv\, \ch^{(*)}_{y'-1}\, \dh_{y+1},
\end{equation}
until the word is standard.

For example, consider the standard word
\begin{equation}
\ch_1\,\ch_2\, \dh_{\frac72}\,
\chs_{\frac{11}2}\, \ch_8\,
\chs_{\frac{19}2}\, \chs_{\frac{21}2}\, \ch_{14}.
\end{equation}
Increasing the excitation of the $\dh_{\frac72}$ by 5, and 
standardising the resulting non-standard word using \eqref{Eq:HalfEquiv},
results in the sequence:
\begin{equation}
\begin{split}
\ch_1\,\ch_2\, (\dh_{\frac{17}2}\,
\chs_{\frac{11}2})\, \ch_8\,
\chs_{\frac{19}2}\, \chs_{\frac{21}2}\, \ch_{14}
&\quad\to\quad
\ch_1\,\ch_2\, \chs_{\frac92}\,
(\dh_{\frac{19}2}\, \ch_8)\,
\chs_{\frac{19}2}\, \chs_{\frac{21}2}\, \ch_{14}\\
&\quad\to\quad
\ch_1\,\ch_2\, \chs_{\frac92}\, \ch_7\,
(\dh_{\frac{21}2}\,
\chs_{\frac{19}2})\, \chs_{\frac{21}2}\, \ch_{14}\\
&\quad\to\quad
\ch_1\,\ch_2\, \chs_{\frac92}\, \ch_7\,
\chs_{\frac{17}2}\,
(\dh_{\frac{23}2}\, \chs_{\frac{21}2})\, \ch_{14}\\
&\quad\to\quad
\ch_1\,\ch_2\, \chs_{\frac92}\, \ch_7\,
\chs_{\frac{17}2}\,
\chs_{\frac{19}2}\, \dh_{\frac{25}2}\, \ch_{14}.
\end{split}
\end{equation}
where in each line, we have used parentheses to indicate the
pair of symbols affected.

To excite a number of integer valleys simultaneously,
we simply add the required excitement to the subscript of
each $\dh_x$ in $\owordh{\hh}$,
and standardise the resulting word by repeatedly
using \eqref{Eq:HalfEquiv}.

We note that, as in the last section,
the excitation of an integer valley corresponding to
an operator $\dh_x$ in an operator word $\osymbolh$ is given by
\eqref{Eq:HALFexcite}, even when the word is non-standard.
This is so because the value of the expression
\eqref{Eq:HALFexcite} is unchanged for operator words obtained
from one another through the equivalence \eqref{Eq:HalfEquiv}.

\subsection{Bijection relating the $\PP^p$ and $\HH^p$ paths}
\label{Sec:BijEquiv}

The bijection between these two path representations of the
$\mathcal{M}(p,2p+1)$ models can now be accomplished by mapping
from the standard operator word corresponding to one,
to an operator word (which may be non-standard) for the other.
The appropriate equivalences,
either \eqref{Eq:RSOSEquiv} or \eqref{Eq:HalfEquiv}, are then used
to obtain the standard operator word, from which the
path of the bijective image is readily read.

For odd integers $a$ and $b$,
with $1\le a<2p$ and $3\le b<2p$,
let $h\in\PP^{p}_{a,b}$ and let
$\hh\in\HH^{p}_{\hat a,\hat b}$ be its image under the
bijection of Section \ref{Sec:BijMap},
where $\hat a=\tfrac12(a-1)$ and $\hat b=\tfrac12(b-1)$.
We claim that if $\osymbol$ is an operator word for $h$,
then $\osymbolh$ is an operator word for $\hh$, where we
obtain $\osymbolh$ from $\osymbol$ by transforming each
operator 
within according to
\begin{equation}\label{Eq:oprsostohk}\cb_{x}  \rightarrow  \ch_{x/2},
\quad \cbs_{y} \rightarrow  \chs_{y/2} ,
\quad \db_{x}  \rightarrow   \dh_{x+1/2}.
\end{equation}

To verify this claim, we show that the paths $h\in\PP^{p}_{a,b}$ and
$\hh\in\HH^{p}_{\hat a,\hat b}$, that correspond to the operator words
$\osymbol$ and $\osymbolh$ respectively, are related through
the map $h\to\hat h$ defined in Section \ref{Sec:BijMap}.

Let the triple $(\hcut,n,\lambda)$ be that corresponding to
$h$, as defined in Section \ref{Sec:BijMap}.
Then $h$ contains $n$ particles and, consequently, $\osymbol$
contains $n$ operators $\db_x$.
Whether $\osymbol$ is standard or non-standard, the excitation
$\lambda_i$ of the particle corresponding to the $i$th operator
$\db_x$, counting from the right, is given by \eqref{Eq:RSOSexcite}.
In the case of a standard word $\osymbol$, the standard word
$\osymbol'=\oword{\hcut}$ of $\hcut$ is obtained from $\osymbol$
(as may be seen from Fig.~\ref{Fig:CCDaction})
simply by reducing the subscript of each term $\cb^{(*)}_z$
by $2\#\{\db\cdots\cb^{(*)}_z\}$
(i.e., by twice the number of particles to its left),
and dropping all operators $\db_x$.
This also holds for non-standard words $\osymbol$,
because $\osymbol'$, obtained in this way, is unchanged under the
equivalence \eqref{Eq:RSOSEquiv}.

We now proceed analogously for the operator word $\osymbolh$,
determining $\mu$ and $\hlac$.
Since $\osymbolh$ has $n$ operators $\db_{x'}$, the half-lattice
path $\hh$ has $n$ integer valleys.
The excitation $\hat\lambda_i$ of the $i$th of these integer valleys,
counting from the right, is given by \eqref{Eq:HALFexcite},
whether $\osymbolh$ is standard or non-standard.
Since the transformation \eqref{Eq:oprsostohk} specifies that $x'=x+1/2$,
we have $\hat\lambda_i=\lambda_i$.
Thereupon, the value of $\mu_i$, the numbering of the $i$th integer
valley of $\hh$ amongst all its valleys, is given by
$\mu_i=\hat\lambda_i+\#\{\dh\cdots\dh_{x'}\}+1=\lambda_i+n-i+1$.
Thus, $\lambda$ and $\mu$ are related by \eqref{Eq:StaggerParts},
as required.

In the case of a standard word $\osymbolh$, the standard word
$\osymbolh'=\owordh{\hlac}$ of $\hlac$ is obtained from $\osymbolh$
(as may be seen from Fig.~\ref{Fig:CCDhaction})
simply by reducing the subscript of each term $\ch^{(*)}_z$
by $\#\{{\dh}\cdots\ch^{(*)}_z\}$
(i.e., by the number of integer valleys to its left),
and dropping all operators $\dh_x$.
This also holds for non-standard words $\osymbolh$, because
$\osymbolh'$, obtained in this way, is unchanged under the
equivalence \eqref{Eq:HalfEquiv}.

Now, in view of the transformation \eqref{Eq:oprsostohk},
we see that the subscripts of the terms in $\osymbol'$
are precisely twice those in $\osymbolh'$.
Thus, the particle-deficient path $\hlac$ is obtained
from $\hcut$ by shrinking by a factor of 2.
Therefore, the combined map
$h\to(\hcut,n,\lambda)\to(\hlac,n,\mu)\to\hat h$
is precisely as specified in Section \ref{Sec:BijMap}.
This completes the verification of this operator
description of the bijection.


To illustrate the description, consider the path
$h\in\mathcal{P}_{1,3}^4$ of Fig.~\ref{TypicalScoringFig}.
Its standard operator word $\oword{h}$ is given in \eqref{Eq:TypicalOword}.
Using the transformation \eqref{Eq:oprsostohk}
and the equivalences \eqref{Eq:HalfEquiv},
we obtain:
\begin{eqnarray}\label{Eq:exbi}
\cb_{0} \: \db_{2} \: \cb_{4}\: \cb_{6}\: \db_{10}\: \cbs_{13}\:
\db_{16}\: \db_{19}\; \cbs_{21}
& \rightarrow &
\ch_{0} \:
\dh_{\frac{5}{2}} \:  \ch_{2} \: \ch_{3}\: \dh_{\frac{21}{2}}\:
\chs_{\frac{13}{2}} \: \dh_{\frac{33}{2}}\: ( \dh_{\frac{39}{2}} \:
\chs_{\frac{21}{2}} )\nonumber \\
 & \rightarrow &
\ch_{0} \:
\dh_{\frac{5}{2}}\: \ch_{2}\: \ch_{3}\: \dh_{\frac{21}{2}}\:
\chs_{\frac{13}{2}} \: ( \dh_{\frac{33}{2}} \: \chs_{\frac{19}{2}} )\:
\dh_{\frac{41}{2}} \nonumber \\ & \rightarrow &
\ch_{0} \:
\dh_{\frac{5}{2}} \: \ch_{2}\: \ch_{3}\: ( \dh_{\frac{21}{2}}\:
\chs_{\frac{13}{2}} )\:  \chs_{\frac{17}{2}} \: \dh_{\frac{35}{2}}\:
\dh_{\frac{41}{2}} \nonumber\\
 & \rightarrow &
\ch_{0} \:
\dh_{\frac{5}{2}} \: \ch_{2}\: \ch_{3}\: \chs_{\frac{11}{2}}\:
( \dh_{\frac{23}{2}}  \chs_{\frac{17}{2}})\: \dh_{\frac{35}{2}}\:
\dh_{\frac{41}{2}} \nonumber\\ & \rightarrow &
\ch_{0} \:
( \dh_{\frac{5}{2}} \: \ch_{2} )\: \ch_{3}\: \chs_{\frac{11}{2}}\:
\chs_{\frac{15}{2}}\: \dh_{\frac{25}{2}}\: \dh_{\frac{35}{2}}\:
\dh_{\frac{41}{2}}\nonumber \\
 & \rightarrow &
\ch_{0} \:
 \ch_{1}\:( \dh_{\frac{7}{2}}  \ch_{3} )\: \chs_{\frac{11}{2}}\:
\chs_{\frac{15}{2}}\: \dh_{\frac{25}{2}}\: \dh_{\frac{35}{2}}\:
\dh_{\frac{41}{2}}\nonumber \\
 & \rightarrow &
\ch_{0} \:
 \ch_{1}\:\ch_2\:( \dh_{\frac{9}{2}}\: \chs_{\frac{11}{2}})\:
\chs_{\frac{15}{2}}\: \dh_{\frac{25}{2}}\: \dh_{\frac{35}{2}}\:
\dh_{\frac{41}{2}}\nonumber \\
 & \rightarrow &
\ch_{0} \:
\ch_{1}\:  \ch_{2}\: \chs_{\frac{9}{2}}\: \dh_{\frac{11}{2}} \:
\chs_{\frac{15}{2}}\: \dh_{\frac{25}{2}}\: \dh_{\frac{35}{2}}\:
\dh_{\frac{41}{2}}.
\nonumber
\end{eqnarray}
Here, we have used parentheses to indicate the
operators reordered by means of (\ref{Eq:HalfEquiv}).
We recognize the final word here to be (\ref{Eq:seq_cced}),
the standard operator word corresponding to the path of Fig.~\ref{che_h}.

To illustrate the inverse mapping, begin with the standard operator
word \eqref{Eq:seq_cced}.
Applying the reverse of the transformations \eqref{Eq:oprsostohk},
and standardising the result using \eqref{Eq:RSOSEquiv} produces
\begin{eqnarray*}
\ch_{0}\: \ch_{1}\: \ch_{2}\: \chs_{\frac{9}{2}}\:
\dh_{\frac{11}{2}}\: \chs_{\frac{15}{2}}\: \dh_{\frac{25}{2}}\:
\dh_{\frac{35}{2}}\: \dh_{\frac{41}{2}} \;
& \rightarrow &
\offinterlineskip
\cb_{0} \: (\cb_{2} \: \cb_{4} \: \cbs_{9}\: \db_{5})\:
\cbs_{15}\: \db_{12} \: \db_{17} \: \db_{20} \\
& \rightarrow &
\cb_{0} \: \db_{2}\: \cb_{4} \: \cb_{6} \:
(\cbs_{11}\: \cbs_{15}\: \db_{12})\: \db_{17} \: \db_{20} \\
& \rightarrow &
\cb_{0} \: \db_{2}\: \cb_{4} \: \cb_{6} \: \db_{10}\: \cbs_{13}\:
(\cbs_{17}\: \db_{17}) \: \db_{20} \\
& \rightarrow &
\cb_{0} \: \db_{2}\: \cb_{4} \: \cb_{6} \:\db_{10}\: \cbs_{13}\:
\db_{16}\: (\cbs_{19}\: \db_{20}) \\
& \rightarrow &
\cb_{0} \:
\db_{2}\: \db_{4}\: \cb_{6}  \:\db_{10}\: \cbs_{13}\: \db_{16}\:
\db_{19} \: \cbs_{21},
\end{eqnarray*}
thus recovering the operator sequence \eqref{Eq:TypicalOword}
of our original path, Fig.~\ref{TypicalScoringFig}.
Here, in each line, we have used parentheses to indicate
a number of applications of \eqref{Eq:RSOSEquiv} which are applied
successively within the included subword to shift the operator
$\db$ completely to the left.


\section{Outlook}
There is a well-known duality between the characters of the $\M(p,p')$
and $\M(p'-p,p')$ models \cite{BMlmp}.
In terms of RSOS paths, this duality involves
interchanging the role of the light and dark bands \cite{FLPW}.
This hints at a description of the $\M(p+1,2p+1)$
states in terms of some sort of dual half-lattice paths.
It turns out that such a description does exist.
Roughly, these are half-lattice paths with half-integer
extremity conditions (instead of integer ones) and with peaks
at half-integer heights. 

To be more precise, let us denote the set of new paths
by $\tilde \HH^p_{\tilde a,\tilde b}$,
where $\tilde a,\tilde b\in\ZZph$.  
This set is defined to contain
all half-lattice paths $\halfh$ that are $(0,p-1/2)$-restricted,
$\tilde b$-tailed, with $\halfh_0=\tilde a$,
and the additional restriction that if
$\halfh_x=\halfh_{x+1}\in\ZZph$, then $\halfh_{x+1/2}=\halfh_x-1/2$.
This additional restriction forces the peaks to occur at non-integer heights
(however, the half-integer initial point ensures that all peaks
occur at integer $x$-positions). 
These paths are weighted
similarly to their dual versions,
using \eqref{Eq:StraightWt} and \eqref{Eq:HalfWt}.
As announced, these new paths describe the states in the $(r,s)$
irreducible module of
$\mathcal{M}(p+1,2 p +1)$ where
\begin{equation}\label{rs_hkd}
r = \tilde a + \frac{1}{2} \qquad \text{and} \qquad s = 2\tilde b.
\end{equation}
Note that, in comparison with the
$\HH^p$ 
paths (see \eqref{rsabh}),
the roles of $\tilde a$ and $\tilde b$ have interchanged in that, here,
$\tilde a$ and $\tilde b$ are related to $r$ and $s$ respectively.
Note also that the vertical
range
is larger for paths in
$\tilde \HH^p_{\tilde a,\tilde b}$ than for those in
$\HH^p_{\hat a,\hat b}$,
being $p-1/2$ and $p-1$ respectively.
However, as previously mentioned, in the latter case, the maximal height
could be augmented to $p-1/2$ without affecting the set of paths since the
constraints on the integrality of the
peaks' heights prevents the extra portion
from being 
reached.
Therefore, both the $\HH^p$ and the $\tilde \HH^p$
paths can be defined in the same strip,
enhancing their duality relationship.

Following the analysis of \cite{PMnpd}, this path representation
leads to new expressions for the $\mathcal{M}(p+1,2 p +1)$
fermionic characters and with a clear particle content.
This will be detailed elsewhere.

Finally, aspects of the present bijection can be turned
into a well-controlled 
exploratory tool for an alternative path description of the
$\M(p,fp+1)$ models on a
$(1/f)$-lattice, 
with special restrictions on the positions of the peaks and valleys.
As already stressed, this is interesting in that it could
lead to novel fermionic forms.
Such a study is left to a future work.

\newtheorem*{acknow}{Acknowledgments}

\begin{acknow}

OBF acknowledges a NSERC student fellowship and thanks the
Institut Henri Poincar\'e  and the Centre \'Emile Borel for its
hospitality in the course of the thematic semester
{\rm Statistical physics, combinatorics and probability:
from discrete to continuous models},
during which part of this work was done. 
This work was  supported by NSERC.

\end{acknow}

\end{document}